\documentclass[%
 reprint,
 amsmath,amssymb,
 aps,
superscriptaddress,
prb,nofootinbib
]{revtex4-2}

\usepackage{graphicx} % Include figure files
\usepackage{bm}% bold math
\usepackage{color}
\usepackage{makecell}
\usepackage{amsmath}
\usepackage{comment}
\usepackage{amssymb}
\usepackage[percent]{overpic}
\usepackage{comment}
\usepackage{enumerate}
\usepackage{float}
\usepackage{xspace}
\usepackage{mathtools}
\usepackage{mathrsfs}
\usepackage{mathptmx}
\usepackage[colorlinks=true,linkcolor=blue,citecolor=blue,urlcolor=blue]{hyperref}
\usepackage{orcidlink}
\usepackage{tikz}
\usepackage{siunitx}
\usepackage[utf8]{inputenc}
\usepackage{nicefrac}
\usepackage[capitalise]{cleveref}

\Crefformat{figure}{#2Fig.~#1#3}
\Crefformat{equation}{#2Eq.~(#1#3)}
\newcommand{\gv}[1]{\ensuremath{\mbox{\boldmath$ #1 $}}}  % bold
\newcommand{\iu}{\textup{i}} % imaginary unit
\newcommand{\ham}{\mathcal{H}}  % Hamiltonian

\newcommand{\wholedm}{Dzyaloshinskii-Moriya }
\newcommand{\e}{\textup{e}} % exponent
\newcommand{\real}[1]{\operatorname{Re}\left(#1\right)}

\newcommand{\smsz}{\ensuremath{\text{S}^z}}
\newcommand{\smJij}{\ensuremath{J_{ij}}\xspace}
\newcommand{\smmu}{\ensuremath{\mu_{\mathrm{s}}}\xspace}

\newcommand{\smB}{\ensuremath{\mathbf{B}}\xspace}
\newcommand{\smA}{\ensuremath{\mathbf{A}}\xspace}

\newcommand{\sms}{\ensuremath{\mathbf{S}}\xspace}
\newcommand{\vampire}{\textsc{vampire} }

\newcommand{\muB}{\ensuremath{\mu_{\mathrm{B}}}\xspace}

\begin{document}

% change the components of pairs of D when tilting it
%similar effect in electronic system under strain

\title{Topological magnon gap engineering in van der Waals CrI$_3$ ferromagnets}

\author{Verena Brehm \orcidlink{0000-0002-7174-1899}}
\affiliation{Center for Quantum Spintronics, Norwegian University of Science and Technology, 7034 Trondheim, Norway}
\author{Pawel Sobieszczyk\orcidlink{0000-0002-1593-1479}} 
\affiliation{Institute of Nuclear Physics Polish Academy of Sciences, Radzikowskiego 152, 31-342 Krakow, Poland}
\author{Jostein N. Kløgetvedt \orcidlink{0009-0009-6759-9920}}
\affiliation{Center for Quantum Spintronics, Norwegian University of Science and Technology, 7034 Trondheim, Norway}
\author{Richard~F.~L.~Evans\orcidlink{0000-0002-2378-8203}}
\affiliation{School of Physics, Engineering and Technology, University of York, York, YO10 5DD, United Kingdom}
\author{Elton J. G. Santos\orcidlink{0000-0001-6065-5787}}
\affiliation{Institute for Condensed Matter Physics and Complex Systems, School of Physics and Astronomy, The University of Edinburgh, Edinburgh EH9 3FD, Untied Kingdom} 
\affiliation{Higgs Centre for Theoretical Physics, The University of Edinburgh,  EH9 3FD, United Kingdom}
\affiliation{Donostia International Physics Center (DIPC), 20018 Donostia-San Sebastián, Basque Country, Spain}
\author{Alireza Qaiumzadeh \orcidlink{https://orcid.org/0000-0003-2412-0296}} 
\affiliation{Center for Quantum Spintronics, Norwegian University of Science and Technology, 7034 Trondheim, Norway} 

\begin{abstract}
The microscopic origin of the topological magnon band gap in CrI$_3$ ferromagnets has been a subject of controversy for years since two main models with distinct characteristics, i.e.,  Dzyaloshinskii-Moriya (DM) and Kitaev, provided possible explanations with different outcome implications.   Here we investigate the angular magnetic field dependence of the magnon gap of CrI$_3$ by elucidating what main contributions play a major role in its generation. We implement stochastic atomistic spin dynamics simulations to compare the impact of these two spin interactions on the magnon spectra. We observe three distinct magnetic field dependencies between these two gap opening mechanisms. First, we demonstrate that the Kitaev-induced magnon gap is influenced by both the direction and amplitude of the applied magnetic field, while the DM-induced gap is solely affected by the magnetic field direction. Second, the position of the Dirac cones within the Kitaev-induced magnon gap shifts in response to changes in the magnetic field direction, whereas they remain unaffected by the magnetic field direction in the DM-induced gap scenario. Third, we find a direct-indirect magnon band-gap transition in the Kitaev model by varying the applied magnetic field direction. These differences may distinguish the origin of topological magnon gaps in CrI$_3$ and other van der Waals magnetic layers. Our findings pave the way for exploration and engineering topological gaps in van der Waals materials.
\end{abstract}

\maketitle

\section{Introduction}
With the experimental demonstration of long-range magnetic order in two-dimensional (2D) van der Waals (vdW) materials \cite{Gong2017intrinsicFMorderInvdWcrystal,FePS3Lee2016,RamanFePs3Wang2016}, 2D magnetic materials have come into focus\cite{Elton_QuantumRescaling2021, Biquadratic_exchange_interactions_in_2D_magnets,TANG20231,Genome22,Coronado23,Yonathan23,Balicas23,Jenkins22,Hicken22,Hicken23,PhysRevB.106.054403,Srini23}. Among them ferromagnetic CrI$_3$ with a honeycomb lattice structure \cite{HuangExpDemoCrI3fm2017} has been attracting intense interest. 
The experimental observation of a gap opening in the Dirac-like magnon spectrum at the K symmetry points of ferromagnetic CrI$_3$ layers \cite{ChenPRX2018} has triggered wide discussions about the microscopic origin of the gap opening. Several proposals suggest that this gap possesses a topological character, originating from either the Dzyaloshinskii-Moriya (DM) \cite{ChenPRX2018,Biquadratic_exchange_interactions_in_2D_magnets,ChangsongDFT,ChenPRX2021,Soenen_Multimodel2023,Theory_of_magnetism_in_the_vdW_magnet_CrI3,milosevicPRB} or Kitaev \cite{Fundamental_Spin_Interactions_Underlying_the_Magnetic_Anisotropy_In_Kitaev_FM,Theory_of_magnetism_in_the_vdW_magnet_CrI3,FirstPrinciplesKitaevXu2018,PerturbTheoryKitaevStavropolos2021,HPTkitaev2020Aguilera,Elliot2021,Soenen_Multimodel2023,XuKitaevNPJ,ChenKee2023_Kitaev} interaction. 
In contrast, alternative theories associate this gap with electron correlations and spin-phonon interactions, implying a non-topological origin \cite{Ke2021,Magnon_Phonon_Delugas2023}.

The existence of a topological magnon gap gives rise to several interesting features and exotic phases in 2D magnetic systems, such as magnon Hall effects, topological magnon and Chern insulator phases, spin Hall effects for Weyl magnons, and magnonic
Floquet topological insulators \cite{reviewTopMagnons, kovalev, Owerre_2017,Li_stackedHallEff,Costa_itinerantFermDescription}. Therefore, it is essential to determine the microscopic origin of the magnon-gap opening and the fundamental interactions that control this topological gap, opening a route to 2D materials with engineered dynamic properties. 

In a recent experiment, it was demonstrated that the Dirac gap at the K points in CrI$_3$ layers remains open and nearly unchanged when an in-plane (IP) magnetic field is applied to induce an IP magnetization configuration \cite{ChenPRX2021}. Apparently, this observation is not compatible with theoretical models featuring next-nearest-neighbor (NNN) DM interactions with an out-of-plane (OOP) DM vector \cite{Owerre2016,Realization_of_Haldane_Kane_Mele_Model_In_System_of_Localized_Spin}. 

Furthermore, recent theoretical studies have shown that it is possible to alter the topological properties of ferromagnetic and antiferromagnetic systems by adjusting the magnetization direction \cite{DMIandKitaevinCrI3analytical,PhysRevB.105.L100402,PhysRevB.101.125111,PhysRevB.106.125103,PhysRevLett.123.237207,kløgetvedt2023tunable,soenenTunableMagTop}. 
Therefore, tuning the magnetic ground state using external magnetic field might be a useful tool to explore the nature of topological magnon bands.

In this paper, we propose that an angular magnetic field dependent analysis of the magnon dispersion relation, more specifically the Dirac gap size and the position of Dirac-like cones, can be used to discriminate between DM and Kitaev interaction mechanisms in CrI$_3$. We combine our analytical linear spin wave theory at zero temperature with numerical results from atomistic spin dynamics simulations at finite but low temperature to study the angular magnetic field dependency of the magnon dispersion. We show that a tilted NNN DM vector may reproduce the results of the recent experiment better than a Kitaev model.

The rest of this paper is structured as follows. In Sec. \ref{sec:model}, we introduce the DM and Kitaev spin model Hamiltonians and our theoretical and numerical methodology. In Sec. \ref{results}, we show the angular-dependent magnon dispersion of these two spin models. In Sec. \ref{sec:proposal}, we suggest relevant observations for examination in future experiments. Finally, we conclude in Sec. \ref{summary}. 

\section{Model}\label{sec:model}
We aim to compare two proposed spin models for a 2D ferromagnetic insulator honeycomb lattice in CrI$_3$.Although a combined model is theoretically feasible in this system, our focus here is on investigating the distinct effects arising from each model.

Magnon branches in both models are anticipated to manifest a topological band gap at K and K' symmetry points of the  Brillouin zone (BZ). 
In this paper, we use a Kitaev model, proposed in Ref. \cite{Fundamental_Spin_Interactions_Underlying_the_Magnetic_Anisotropy_In_Kitaev_FM}, and compare it width a DM model, proposed in Ref. \cite{Biquadratic_exchange_interactions_in_2D_magnets} to describe the spin dynamics in CrI$_3$. 
For the Kitaev model, the spin interaction Hamiltonian includes the bond-directional anisotropy given by \cite{Fundamental_Spin_Interactions_Underlying_the_Magnetic_Anisotropy_In_Kitaev_FM}
\begin{equation}
    \begin{aligned} \label{eq:H_Kitaev}
    \mathcal{H}_{\kappa} = &- J_\kappa\sum_{\langle i,j\rangle} \sms_i \cdot \sms_j  - D_z \sum_i \left(\smsz_i\right)^2 - \smmu h_0 \sum_i \smB \cdot \sms_i  \\&- \kappa \sum_{\langle i,j\rangle \in \eta } \text{S}_i^\eta \text{S}_j^\eta,
\end{aligned}
\end{equation}
while the spin-interaction Hamiltonian for the DM model reads \cite{Biquadratic_exchange_interactions_in_2D_magnets},
\begin{equation}
\begin{aligned}\label{eq:H_DM}
\mathscr{H}_\text{DM} = &- \sum_{i<j} \big(\smJij \sms_i \cdot \sms_j + \lambda_{ij} \smsz_i \smsz_j \big) - D_z \sum_i \left(\smsz_i\right)^2  \\& - \smmu h_0 \sum_i \smB \cdot \sms_i   - K_\text{bq}\sum_{\langle i,j\rangle}\left(\sms_i \cdot \sms_j \right)^2 \\ &- \sum_{\langle\langle i,j\rangle\rangle} \smA_{ij} \cdot \sms_i \times \sms_j.
\end{aligned}\end{equation}
In the above Hamiltonians, $\sms_i$ is a unit vector that carries the spin-moment direction at site $i$ \cite{vampire}, \smB is the direction of the external magnetic field with strength $h_0$, $\mu_s$ is the atomic magnetic moment, and $D_z>0$ is the OOP easy-axis magnetic anisotropy along the $z$ direction. 
The symbols $\langle i, j \rangle$ and $\langle \langle i, j \rangle \rangle$ represent the sums over first-nearest neighbor (NN) and NNN sites, respectively. The direction of the magnetic field determines the ground-state magnetization direction when its amplitude is larger than a critical value, dictated by the magnetic anisotropy.
In the Kitaev model, \cref{eq:H_Kitaev}, $J_{\kappa}$ represents the NN isotropic Heisenberg exchange interaction, and $\kappa$ denotes the NN bond-$\eta$-dependent Kitaev interaction strength. In the DM model, \cref{eq:H_DM}, the bilinear Heisenberg exchange interaction is split into the isotropic $J_{ij}$ and anisotropic $\lambda_{ij}$ terms, and the sum runs over up to the third NNs. In this spin Hamiltonian, $K_{\rm{bq}}$ is the strength of the NN biquadratic exchange interaction, which renormalizes the isotropic Heisenberg exchange interactions, see \cref{A:EqAbbreviations2}, and $\smA^{\text{NNN}}_{ij} = \nu_{ij} (A_{x}^{\text{NNN}} \gv{\hat{x}} + A_y^{\text{NNN}} \gv{\hat{y}} + A_z^{\text{NNN}} \gv{\hat{z}})=\nu_{ij} \left(A, \theta_{\rm DM}, \varphi_{\rm DM}\right)$ is the DM vector \cite{Soenen_Multimodel2023,milosevicPRB}, with $\nu_{ij}=-\nu_{ji}=\pm 1$,  $A=|\smA^{\text{NNN}}_{ij}|$, and
$\theta_{\rm DM}$ ($\varphi_{\rm DM}$) is the polar (azimuthal) angle. For simplicity, we set $\varphi_{\rm DM} = 0$, which corresponds to the IP $x$ direction, while the ground-state magnetization direction can vary in different directions with respect to the DM vector. 
The average magnetization direction of the ground state, $\mathbf{m}={N}^{-1}\sum_i^N \sms_i$, is a vector $\mathbf{m} = (1,\theta_m,\varphi_m)$ normalized to the unit length that can be controlled by an external magnetic field. We define the relative angle between the DM vector and the magnetization direction as $\theta = \theta_{\rm DM} - \theta_m$.

\begin{figure}
\includegraphics[width=\linewidth]{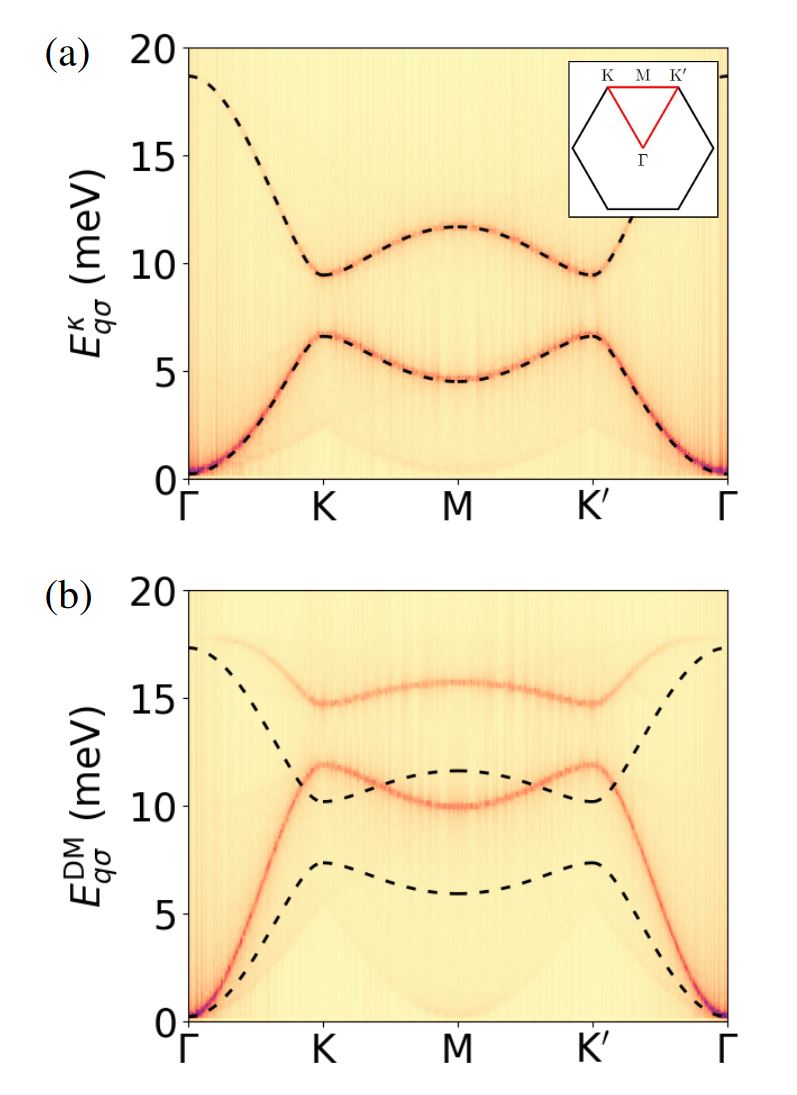}
\caption{Magnon dispersion of single-layer CrI$_3$ in the out-of-plane magnetization configuration in the absence of magnetic field as obtained from atomistic spin simulations (color map) and linear spin wave theory (black dashed lines). The parameters for (a) the Kitaev model and (b) for the DM model can be found in \Cref{A:parameters}. Note that in the \wholedm (DM) model, only first-nearest-neighbor (NN) exchange is included in the analytical model, while up to third NNs are included in the atomistic simulation. This leads to a stretching of the low-energy band and a compression of the high-energy band \cite{Biquadratic_exchange_interactions_in_2D_magnets} but does not affect the size and position of the topological Dirac gap at the K and K' points.}
\label{fig:dispersionComparison}
\end{figure}

\subsection{Magnon dispersion from linear spin-wave theory}
We analytically calculate the noninteracting magnon dispersion for an arbitrary ground-state magnetization direction, determined by an external magnetic field, in both Kitaev and DM models. To compute the magnon dispersion relations, we use the standard linear spin-wave theory by applying the Holstein-Primakoff transformation \cite{Holstein_Primakoff} at zero temperature. 
The noninteracting magnon Hamiltonian in the second quantized representation reads,
\begin{align}
\mathscr{H}_{\kappa (\rm{DM })} = 
\sum_{\mathbf{q},\sigma} E^{\kappa (\rm{DM })}_{\mathbf{q}\sigma} a_{\mathbf{q},\sigma}^\dagger a_{\mathbf{q},\sigma},   
\end{align}
where $a_{\mathbf{q},\sigma} (a^{\dagger}_{\mathbf{q},\sigma})$ is the bosonic annihilation (creation) operator for acoustic-like, $\sigma=-$, and optical-like, $\sigma=+$, magnon modes with eigen energy $E^{\kappa (\rm{DM})}_{\mathbf{q}\sigma}$.  
In the Kitaev (DM) model, a topological Dirac magnon gap can be opened at the K points depending on the tilting angle of the magnetization direction $\theta_m$ (or the relative angle between the DM vector and magnetization $\theta$ respectively), $\Delta^{\kappa (\rm{DM })}_{\rm{K}}= E^{\kappa (\rm{DM })}_{\mathbf{q}=\rm{K},\sigma=+}-E^{\kappa(\rm{DM })}_{\mathbf{q}=\rm{K},\sigma=-}$. 

In the OOP magnetization configuration, the DM model with an OOP DM vector leads to a topological magnon gap at the K points which is linearly proportional to the OOP DM strength, $\Delta^{\rm{DM }}_{\rm{K}}(\theta=0) \approx A_z^{\rm{NNN}}$,
while in the Kitaev model, the K-point gap depends on different spin interactions and, more importantly, on the external OOP magnetic field  
$ \Delta^{\kappa}_{\rm{K}}(\theta=0) \approx t_0-\sqrt{t_0^2 -\kappa^2\left(\frac{3}{2}\right)^2}$, with $t_0=(9J_\kappa + 3\kappa)/2+\mu_s h_0$, with $\kappa,\mu_s h_0,J_\kappa \gg D_z$. From these topological magnon gap expressions, it is evident that in the OOP magnetic state, unlike the DM-induced topological gap, the Kitaev-induced topological gap can be tuned by varying the strength of the OOP magnetic field \cite{DMIandKitaevinCrI3analytical}. In general, both the OOP easy-axis magnetic anisotropy and magnetic field strengths modify the topological gap value in the Kitaev model, see \Cref{A:analyticalCalc}.

Since the magnetic unit cell in a honeycomb lattice has two spins, there are two magnon branches in the CrI$_3$ single layer, as shown in \Cref{fig:dispersionComparison}. The black dotted curves in \Cref{fig:dispersionComparison} show the analytical magnon dispersion for (a) the Kitaev and (b) the DM model respectively.
Note that in the analytical calculations involving the DM Hamiltonian, \Cref{eq:H_DM}, we solely consider the isotropic NN exchange interaction in the first term of the Hamiltonian. Including all NNs in the simulation of the atomistic spin model leads to a stretching of the acoustic branch and compression of the optic branch \cite{Biquadratic_exchange_interactions_in_2D_magnets}, as shown by the red lines.

Since the magnon dispersion expressions for arbitrary ground-state magnetization directions are lengthy, we only show a graphical representation of the results in the main text and refer to \cref{A:analyticalCalc} for the analytic expressions.

\subsection{Magnon dispersion from atomistic spin dynamics simulations}
The dynamic of spins within our two models is simulated using the \vampire software package \cite{vampireURL,vampire} that solves the stochastic Landau–Lifshitz-Gilbert (sLLG) equation, applied at the atomistic level~\cite{vampire,EllisLTP2015}, numerically. The sLLG equation reads
\begin{equation}
\label{eqn:LLG}
\frac{\partial \mathbf{S}_i}{\partial t} = -\frac{\gamma}{1 + \alpha^2} \left[\mathbf{S}_i \times \mathbf{B}^{\rm \kappa (DM)}_i + \alpha \mathbf{S}_i\times \left(\mathbf{S}_i \times  \mathbf{B}^{\rm \kappa (DM)}_i \right)\right],
\end{equation}
where $\gamma$ is the electron gyromagnetic ratio and $\alpha$ is the Gilbert damping constant. The effective magnetic field for the Kitaev (DM) model $\mathbf{B}^{\rm \kappa (DM)}_i = - {\mu_s}^{-1}{\partial \mathcal{H}_{\rm \kappa (DM)}}/{\partial \sms_i} + \mathbf{\xi}^{\rm (th)}_i$, consists of a deterministic contribution from the corresponding spin-interaction Hamiltonian, the first term, and a stochastic thermal field, the second term. The latter introduces temperature to the system and is modeled by an uncorrelated Gaussian thermal noise that obeys,
\begin{subequations}
\begin{align}\label{eq:noise}
     &\langle\mathbf{\xi}^{\rm (th)}_i(t)\rangle =0,  \\ 
    & \langle\xi_{i,m}^{\rm (th)}(t) \xi_{j,n}^{\rm (th)}(t')\rangle = 2 \alpha k_\mathrm{B}T \gamma^{-1}\mu_s^{-1}\delta_{ij}\delta_{mn}\delta(t - t'), 
\end{align}
\end{subequations}
where $m,n = \{ x,y,z \}$ represent spatial components and $k_\mathrm{B}$ is the Boltzmann constant.  

To simulate spin dynamics in single-layer CrI$_3$, we consider a honeycomb lattice with a size of 300 $\times$ 300 unit cells, $\sim 1.8 \times 10^5$ spins, at low but finite temperature. 
A preconditioning adaptive-step Monte Carlo simulation is employed to achieve thermal equilibrium within the system~\cite{Alzate-CardonaJPCM2019}. To ensure rapid convergence, the damping constant $\alpha$ is set to $1$ during the preconditioning. In the next step we integrate the stochastic Langevin \Cref{eqn:LLG} dynamically using a stochastic Heun method \cite{vampire} over \SI{30}{ps}, with a time step of \SI{5}{fs}, and damping $\alpha=0.01$. This choice of time step, along with low damping, is sufficient to obtain accurate magnon spectra \cite{Etz_2015,Barker2016}. The magnon spectra for a path in momentum space can be derived by computing the fast Fourier transform of spatially and temporally dependent spin moment directions. For more details, see \Cref{E:NumericalCalculationRD}.

At low temperatures, the density of thermal magnons is low, and nonlinear magnon interactions are comparatively weak. As temperatures increase, these nonlinear interactions may give rise to topological phase transitions \cite{DiracGapClosingHighTcLu2021}. 

The magnon spectra obtained from the atomistic spin dynamics simulations at low temperatures closely align with the predictions of the linear spin-wave theory at zero temperature, as shown for the OOP magnetization configuration in \Cref{fig:dispersionComparison} and various orientations of the ground-state magnetization in \Cref{DMIappendix,KitaevAppendix}.

\section{Engineering of the magnon dispersion relation}\label{results}
In this section, we employ the numerical and analytical methods outlined in the previous section to investigate the impact of the magnetization direction on the magnon spectra for the two proposed spin models of CrI$_3$.

\begin{figure}[tb]
\includegraphics[width=\linewidth]{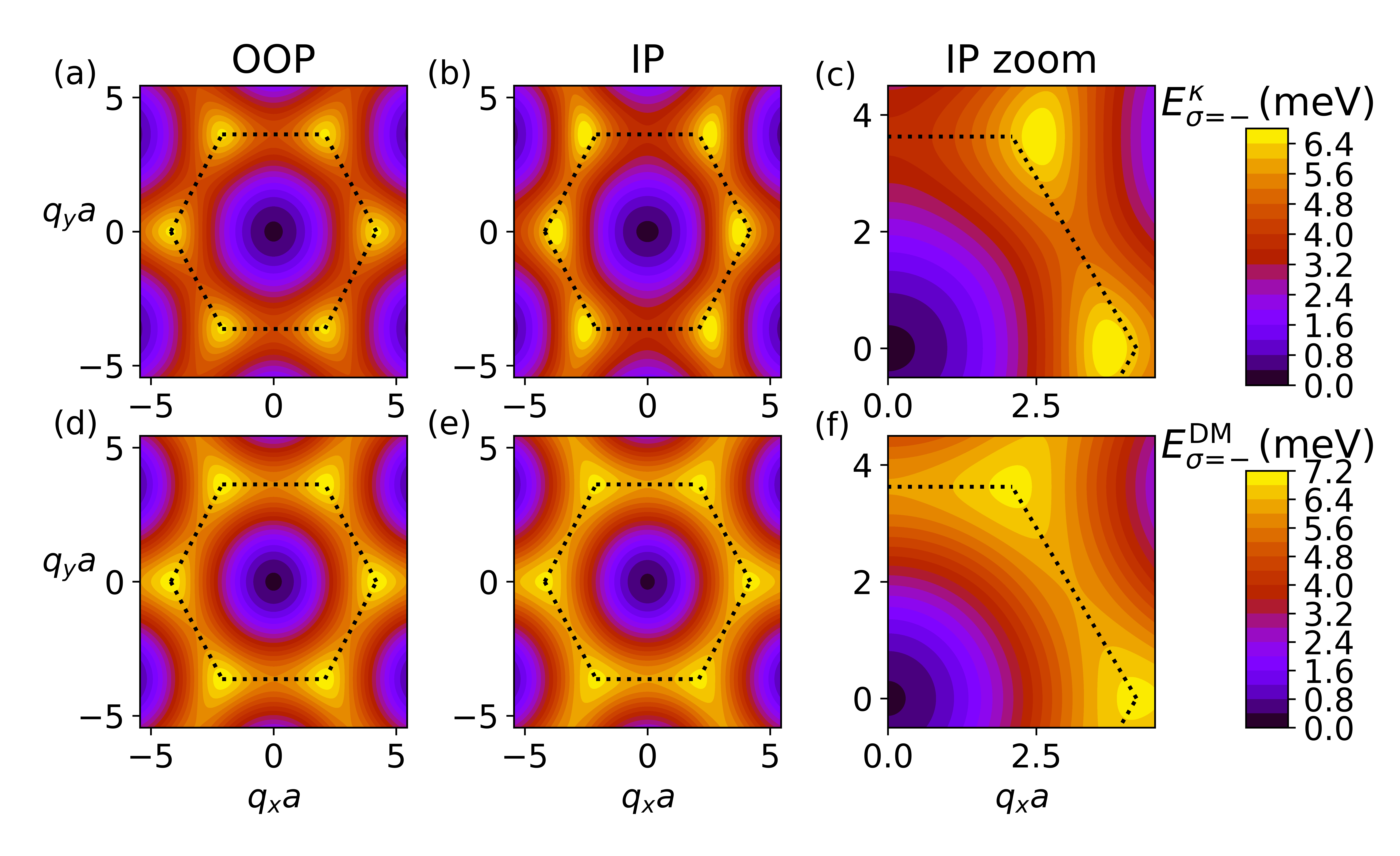}
    \caption{Analytic comparison of the positions of the Dirac-like cones in the magnon dispersion relation at out-of-plane (OOP, left column) and in-plane (IP, right columns) orientation with constant-energy cuts. In the Kitaev ($\kappa$) model, the Dirac-like cones are at the K-points in (a) the OOP configuration but migrate in (b) and (c) the IP configuration. In contrast, (d-f) in the \wholedm (DM) model with a tilted DM vector the cones remain at the K points.  Here, the low-energy band $\sigma=-$ is shown and the color map scales from blue (low energy) to red and yellow (high energy). The magnetic field strength is \SI{4.5}{T} \cite{ChenPRX2021}.}
    \label{fig:comparisonExtremaIP_OOP}
\end{figure}

\subsection{Migration of the Dirac gap in the Kitaev model} \label{sec:Kitaev}
The previously introduced Kitaev model in \Cref{eq:H_Kitaev} successfully replicates the experimentally observed magnon band gap at the $\Gamma$ point and Dirac points, as reported in Ref. \cite{ChenPRX2021}. This is achieved by employing the following spin parameters: $J=\SI{0.55}{meV}$, $\kappa=\SI{4.5}{meV}$, and $D_z=\SI{0.1}{meV}$. 

The Kitaev interaction $\kappa$ opens a Dirac gap at the K and K' points as presented in \Cref{fig:dispersionComparison}(a), where the magnon dispersion is shown for an OOP magnetization ground state. In this figure, we compare numerical results from atomistic spin simulations at low temperature, indicated by the color map, with analytical results from linear spin-wave theory at zero temperature, drawn with the black dashed line. We find excellent agreement between both methods and that the $\Gamma$ and K point gaps are comparable with the experimental values ~\cite{ChenPRX2018,ChenPRX2021}.  

In \Cref{fig:comparisonExtremaIP_OOP}(a), we present constant-energy cuts of the magnon dispersion for the OOP magnetization direction. Here, the Dirac-like cones sit at the K and K' points because of $C_{3v}$ symmetry \cite{PhysRevB.75.155424,kogan2012dirac}. 
However, when rotating the magnetization direction to the IP configuration, e.g., along the $x$-direction, although the Dirac gaps remain open, the Dirac cones are displaced from the high-symmetry K and K' points, see Figs.~\ref{fig:comparisonExtremaIP_OOP}(b) and \ref{fig:comparisonExtremaIP_OOP}(c). 
The two Dirac cones along the $q_y=0$ line migrate toward the center of the BZ, while the other four Dirac cones, with finite $q_y$, move outward from the original BZ. Hence, the BZ is squeezed along the $y$ direction when spins are along the $x$ direction in this model. If the IP magnetization direction is set along the $y$ direction, the BZ is squeezed along the $x$ direction (not shown).
This displacement of the Dirac cones should be distinguished from recently reported intensity widening \cite{ChenPRX2021} and Dirac nodal lines \cite{ChenPRX2021, Elliot2021,NodalLines}. 

While the magnetic field direction moves the Dirac points, the magnetic field strength modifies the Dirac gap size, which will be discussed below.

\subsection{Tuning of the Dirac gap size in the DM model} \label{sec:DMI}
To investigate magnon dispersion relations in the DM model in \Cref{eq:H_DM}, we use the spin interaction parameters reported in Ref. \cite{Biquadratic_exchange_interactions_in_2D_magnets}, which are listed in \Cref{A:parameters}.
However, since a magnon gap is only opened at Dirac points if the ground-state magnetization direction has a finite projection on the NNN DM vector, i.e., $\mathbf{A}\cdot \mathbf{m} \neq 0$ or $\theta \neq 90^\circ$, we argue that the NNN DM vector must be tilted. Only when the NNN DM vector is tilted can a finite magnon gap be opened in both OOP and IP magnetic configurations as observed in recent magnon dispersion measurements in Ref.~\cite{ChenPRX2021}.

In a pristine magnetic layer with honeycomb lattice structure, the intrinsic NNN DM vector is perpendicular to the plane \cite{Ultrafast_control_spin_honeycomb} by the constraints of symmetry. However, we argue that in realistic layered vdW magnetic materials, such as CrI$_3$, this intrinsic DM vector might be tilted by reducing the lattice symmetry due to various reasons. First, a single layer of these magnetic materials consists of several nonmagnetic atomic layers that break the mirror and inversion symmetries \cite{AlirezaDFT,ChangsongDFT}.
In the case of monolayer CrI$_3$, the magnetic Cr ions arrange themselves in a honeycomb lattice, where each Cr atom is surrounded by six I atoms, creating a distorted octahedral structure through edge sharing.
Second, strain can induce lattice distortion and/or inversion symmetry breaking \cite{Edstrom2022,Basak_2023,Ren2023,AlirezaDFT}. The strain can be externally applied or caused by growing CrI$_3$ on a substrate. Third, it is worth noting that magnon dispersions have, thus far, been measured exclusively in multilayered vdW systems and not in a truly single magnetic layer. This may lead to the deviation of the NNN DM vector from the OOP direction by inversion symmetry breaking \cite{Soenen_Multimodel2023}.

\begin{figure}[tb]
    \centering
    \includegraphics[width=\linewidth]{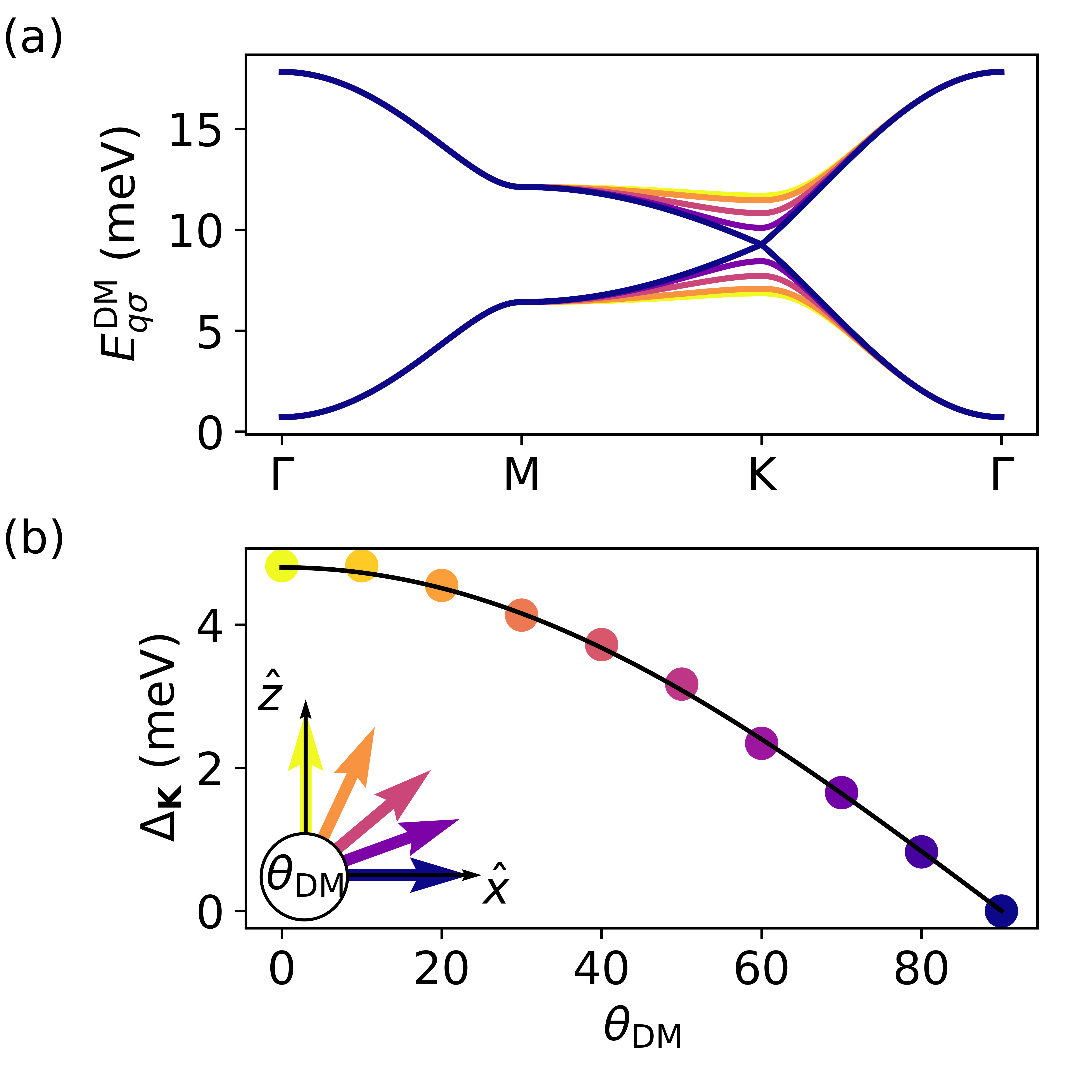}
    \caption{ Impact of \wholedm (DM) interaction on the magnon dispersion relation in the out-of-plane magnetization configuration. (a) Tilting the DM vector only impacts the edge of the Brillouin zone (BZ). The gap size is read out and presented below. (b) When the relative angle $\theta_{\rm{DM}}$ between the DM vector and the magnetization direction increases from parallel $\theta_{\rm{DM}}=0^\circ$ (yellow) to orthogonal $\theta_{\rm{DM}}=90^\circ$ (blue), the gap at the K-symmetry point $\Delta_{\rm{K}}$ closes. The numerical results shown as points agree with the analytical result $\Delta^{\rm{DM}}_{\rm{K}}(\theta_{\rm{DM}}) = \Delta^{\rm{DM}}_{\rm{K}}(0) \cos\theta_{\rm{DM}}$, drawn with the black line, where $\Delta^{\rm{DM}}_{\rm{K}}(0) =9\sqrt{3}A$.}
    \label{fig:DMItiltingAngle}
\end{figure}

In \Cref{fig:DMItiltingAngle}, we show how tilting the DM vector changes the value of the magnon gap at the K point when the magnetic ground state is OOP $\theta_m=0$. It is evident from \Cref{fig:DMItiltingAngle}(a) that the direction of the DM vector only has an impact at the edges of the BZ. The analytical linear spin-wave theory is in perfect agreement with the magnon dispersion computed numerically, see \Cref{DMIappendix}. The size of the magnon gap at the K point, depending on the DM angle $\Delta_{\rm{K}}(\theta_{\rm{DM}})$, is read out and presented in \Cref{fig:DMItiltingAngle}(b) with colored points for the numerical solution and the solid black line for the analytical solution. At $\theta_{\rm{DM}}=0^\circ$ (yellow), where the magnetization direction and the DM vector are parallel, the magnon band gap at the high-symmetry K and K' points is maximal. With increasing angle between the magnetization and the DM vector, the gap reduces, until at $\theta_{\rm{DM}}=90^\circ$ (blue), where the magnetization direction and the DM vector are orthogonal and the magnon band gap at the high-symmetry K and K' points is closed.

Assuming a DM strength of $A \approx \SI{0.31}{meV}$ as reported in Ref. \cite{Biquadratic_exchange_interactions_in_2D_magnets}, we find that a DM tilting angle of $\theta_{\rm{DM}}=54^\circ$ reproduces the reported magnon band gap at the K point $\Delta_{K} \approx \SI{2.8}{meV}$, in Ref.~\cite{ChenPRX2021}. 
It should be stressed that only the magnitude of the DM interaction and the relative orientation between the DM vector and the magnetization direction is relevant for the size of the magnon gap at Dirac points. 
Through the application of an external magnetic field and the rotation of the CrI$_3$ sample, it becomes feasible to experimentally engineer the DM-induced topological band gap.
In the DM model, in contrast to the Kitaev model, the Dirac cones remain at the K and K' points for both OOP and IP magnetization, see Figs.~\ref{fig:comparisonExtremaIP_OOP}(d)-\ref{fig:comparisonExtremaIP_OOP}(f).

\section{Discussion and Proposal} \label{sec:proposal}
In \Cref{fig:comparisonExtremaIP_OOP}, we present a comparative analysis of constant-energy cuts of magnon dispersion within the BZ for IP and OOP magnetic geometries in both Kitaev [Figs.~\ref{fig:comparisonExtremaIP_OOP}(a)-\ref{fig:comparisonExtremaIP_OOP}(c)] and DM [Figs.~\ref{fig:comparisonExtremaIP_OOP}(d)-\ref{fig:comparisonExtremaIP_OOP}(f)] models. 
In all subplots, the BZ of a hexagonal lattice is depicted for comparison. For an OOP magnetization ground state, Figs.~\ref{fig:comparisonExtremaIP_OOP}(a) and \ref{fig:comparisonExtremaIP_OOP}(d), all Dirac-like cones are at the K and K' points in both Kitaev and DM models. However, with an IP magnetization ground state, Figs.~\ref{fig:comparisonExtremaIP_OOP}(b), \ref{fig:comparisonExtremaIP_OOP}(c), \ref{fig:comparisonExtremaIP_OOP}(e), and \ref{fig:comparisonExtremaIP_OOP}(f), although the Dirac-like cones remain at the K and K' points in the DM model, they shift in the Kitaev model.
    \begin{figure}
        \includegraphics[width=\linewidth]{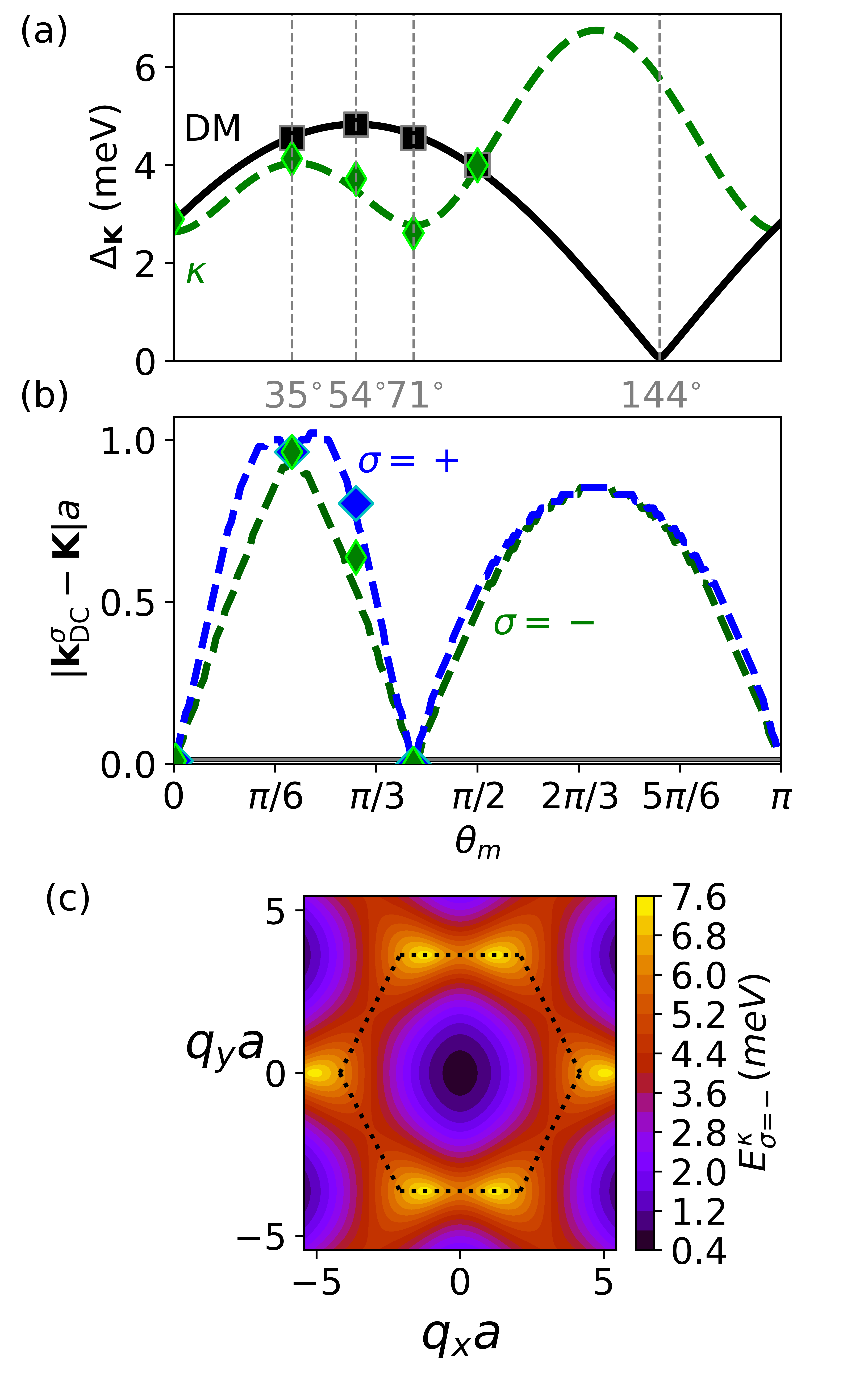}\caption{Impact of the tilting angle $\theta_m$ on dispersion, from out-of-plane (OOP, $\theta_m=0$) over in-plane (IP, $\theta_m=\pi/2$) to negative OOP ($\theta_m=\pi$). Solid lines are obtained analytically within linear spin wave theory and in agreement with numerical data, obtained from atomistic simulations, for selected angles, shown with squares for the \wholedm (DM) model and diamonds for the Kitaev model. (a) Gap at the K point $\Delta_{\rm{K}}$ for the Kitaev (green dashed) and DM model (black solid) models dependent on the magnetization direction $\theta_m$. The DM direction is fixed at $54^\circ$.  (b) Displacement of the Dirac-like cones at $\mathbf{k}^\sigma_{\rm{DC}}$ from the K point in the acoustic ($\sigma=-$) and optic ($\sigma=+$) branches for the Kitaev (green dashed lines) and DM (black and gray solid lines, both remaining zero) models. Characteristic angles indicated by vertical lines. (c) Constant-energy cut in the Kitaev model at $\theta_m = 35^\circ$ showing the migration of the Dirac gap. For tilting the magnetization, a magnetic field of \SI{4.5}{T} is applied \cite{ChenPRX2021}.}
        \label{fig:Kitaev_vs_DMI_gap_and_shift}
    \end{figure}

We propose that two quantities must be experimentally investigated under variation of the ground-state magnetization direction $\theta_m$: first, the size of the band gap at the K and K' points $\Delta_{\rm{K}}$, and second, the position of the Dirac-like cones $\mathbf{k}^\sigma_{\rm{DC}}$ with respect to the K and K' points, $(|\mathbf{k}^\sigma_{\rm{DC}} - \rm{\mathbf{K}}|a)$, for each magnon branch $\sigma$. 
    
In \Cref{fig:Kitaev_vs_DMI_gap_and_shift}(a), we compare the angular dependence $\theta_m$ of the topological magnon gap at the K point $\Delta_{\rm{K}}(\theta_m)$ for the DM model with a tilted DM vector (black solid line) and the Kitaev model (green dashed line). 
In our model, the DM vector is tilted by $54^\circ$. Consequently, the Dirac gap remains open for both OOP with $\theta_m=0$ and IP with $\theta_m=90^\circ$ magnetization directions. As previously discussed, the Dirac gap closing is influenced by the relative angle between the NNN DM vector and the magnetization direction. 
Although our spin-model parameters used in both  Kitaev and the DM models reproduce a topological magnon gap at IP and OOP magnetization directions comparable with the recent experimental data \cite{ChenPRX2021}, the angular dependence of them is quite different. In the Kitaev model, the magnon gap at the K and K' points, $\Delta_{\rm{K}}$, varies with the external magnetic field direction but never closes.
On the contrary, in the DM model with tilted DM vector, the topological gap is largest when the ground state magnetization and DM vector are parallel and closes when magnetization and DM vector are orthogonal. In \Cref{fig:Kitaev_vs_DMI_gap_and_shift}(a) four angles are indicated with vertical dashed lines that are characteristic for our model.

As we mentioned before, the magnon Dirac-like cones in the DM model remain at the K and K' points for all magnetization directions. In contrast, in the Kitaev model, the Dirac cones are displaced by varying the magnetization direction. This can be explained in terms of threefold rotational symmetry about the $z$ axis, the $C_{3z}$ spin point group symmetry \cite{Corticelli2022}. In the OOP configuration, both the Kitaev and DM interactions preserve this symmetry, and the Dirac points remain at the K and K' points. However, with IP magnetization, the Kitaev interaction breaks $C_{3z}$ symmetry, which allows the Dirac points to move away from the K and K' points. On the other hand, in the DM model $C_{3z}$ symmetry is preserved even when there is a finite angle between the DM vector and the magnetization direction, fixing the Dirac points to the K and K' points \cite{Corticelli2022}. 

Furthermore, there are several direct-indirect band gap transitions in the magnon spectra.    
In \Cref{fig:Kitaev_vs_DMI_gap_and_shift}(b), the position of the Dirac cones $\mathbf{k}^\sigma_{\rm{DC}}$ with respect to K and K' points are shown for each branch $\sigma$ in both models. In the DM model (gray and black lines), the cones remain at the K point, and the magnon band gap is always direct. However, in the Kitaev model, shown by dashed green and blue lines, we observe a significant displacement of cones with maximal displacement around $\theta_m=35^\circ$. In addition, as shown by a different amplitude of displacement for the two branches, the band gap is indirect where the blue ($\sigma=+$) and green ($\sigma=-$) branches overlap. To illustrate the displacement of the Dirac gap in the Kitaev model, a constant-energy cut is presented in \Cref{fig:Kitaev_vs_DMI_gap_and_shift}(c) for $\theta_m=35^\circ$, clearly showing a migration of the Dirac cones away from the K and K' points.  In Figs.~\ref{fig:Kitaev_vs_DMI_gap_and_shift}(a) and \ref{fig:Kitaev_vs_DMI_gap_and_shift}(b), we find good agreement between our analytical solution, shown with solid lines, and the numerical results at selected angles, shown with markers. Our atomistic spin simulations reproduce all crucial features: the changing topological magnon gap size for both models as well as a shift in the Dirac cone positions and a direct-indirect gap transition in the Kitaev model.

We have numerically tested a combined model that includes both Kitaev and DM interactions and find that the two mechanisms cooperate in the Dirac gap opening, but only the Kitaev interaction determines the Dirac cone position.

To test our predictions -- that the dominant mechanism of opening the topological magnon gap at the Dirac points of CrI$_3$ is the DM interaction and the Kitaev interaction is negligible -- we propose measuring the magnon dispersion at varying external magnetic field angles. By comparing the topological gap values and the shift of the Dirac-like cones, we can determine whether the DM or Kitaev interaction is the primary mechanisms responsible for the topological gap opening at the Dirac points.

So far, we have examined only the effect of the magnetic field direction, equivalent to the magnetization direction, on the magnon dispersion. However, we have already mentioned that the size of the magnon gap at the Dirac points in the Kitaev model also depends on the amplitude of the magnetic field, as also predicted in Ref. \cite{DMIandKitaevinCrI3analytical}. Thus, we suggest the analysis of the the gap value in the presence of an OOP magnetic field as an another possible experimental study. For example, applying an external magnetic field of \SI{9}{T} in the OOP direction, the gap would decrease by \SI{0.36}{meV} ($\sim$13\%) compared with the case without the magnetic field. It is worth pointing out that the magnetic field strength does not have an impact on the migration of the Dirac cones, and thus the characteristic external magnetic field angles that are indicated in \Cref{fig:Kitaev_vs_DMI_gap_and_shift}(a) remain unchanged. %both of this was verified numerically

\section{Outlook}\label{summary}
We have reexamined the unresolved issue regarding the microscopic origin of the topological magnon band gap observed experimentally at the high-symmetry K and K' points of ferromagnetic CrI$_3$ single layers. This investigation involves a comparison of the angular dependence of the Dirac magnon gap size and its position in the DM model with the Kitaev model.
We have shown that in the Kitaev approach, the size and position of the Dirac points are related to the amplitude and orientation of an applied magnetic field. In contrast, in the DM model, the magnon band gap is related to the angle between the ground-state magnetization direction and the NNN-DM vector. We propose that a tilted DM vector may explain recent magnon dispersion measurements in CrI$_3$ layers \cite{Magnetic_Field_Effect_on_Topological_Spin_Excitations} and motivate further experimental work to engineer intrinsic nontrivial interactions. Based on our findings, we suggest that experimentally exploring the angular dependence of the magnon gap will not only serve as a valuable route to investigate its microscopic origin but also offer a clear pathway to manipulate and tailor topological gaps on-demand accordingly with the target applications. 

\section*{acknowledgments}
V.B. acknowledges L. Chen and P. Dai for helpful discussions.
This project has been supported by the Norwegian Financial Mechanism Project No. 2019/34/H/ST3/00515, ``2Dtronics''.
P.S. was supported by the Polish National Science Centre with Grant Miniatura No. 2019/03/X/ST3/01968.
A.Q. was partially supported by the Research Council of Norway through its Centers of Excellence funding scheme, Project No. 262633, ``QuSpin''. EJGS acknowledges computational resources through the CIRRUS Tier-2 HPC Service (ec131 Cirrus Project) at EPCC (http://www.cirrus.ac.uk) funded by the University of Edinburgh and EPSRC (EP/P020267/1). EJGS acknowledges the Edinburgh-Rice Strategic Collaboration Awards and the EPSRC Open Fellowship (EP/T021578/1) for funding support.

\onecolumngrid
\appendix 

\renewcommand{\thefigure}{S\arabic{figure}}

\renewcommand{\thetable}{S\arabic{table}}

\section{Atomistic simulation parameters} \label{A:parameters}
The parameters used in the atomistic spin simulation for the Kitaev model as inspired by Ref. \cite{Fundamental_Spin_Interactions_Underlying_the_Magnetic_Anisotropy_In_Kitaev_FM} are shown in \cref{tab:parametersKitaevmodel}, and for the DM model as taken from Ref. \cite{Biquadratic_exchange_interactions_in_2D_magnets} in \cref{tab:parametersDMImodel}.

\begin{table}[h]
%\caption{Atomistic spin simulation parameters for the two models}
\begin{minipage}[t]{.45\linewidth}
      \centering
        \caption{Atomistic spin simulation parameters for the Kitaev model \label{tab:parametersKitaevmodel}}
        \begin{ruledtabular}
        \begin{tabular}{c c c} % centered columns (4 columns)
        Parameter & Symbol & Value \\
        \hline % inserts single horizontal line
         g-factor & $g_{e}$      & 2 \\ \hline
         Spin  & $S$ & $3/2$\\ \hline
         Magnetic moment & $\smmu=g_{e}S\muB$      & 3 $\muB$  \\ \hline
         Isotropic bilinear exchange & $ J_\kappa$   & \SI{0.55}{meV}  \\ \hline
        Easy axis anisotropy & $D_z$ & \SI{0.10882}{meV} \\  \hline
        Kitaev strength & $\kappa$ & \SI{4.5}{meV}
        \end{tabular}
        \end{ruledtabular}
        \end{minipage} \hspace{0.5cm}
        \begin{minipage}[t]{.45\linewidth}
      \caption{Atomistic spin simulation parameters for the DM model\label{tab:parametersDMImodel}}
      \centering
        \begin{ruledtabular}
        \begin{tabular}{c c c} % centered columns (4 columns)
        Parameter & Symbol & Value \\
        \hline % inserts single horizontal line
         g-factor & $g_{e}$      & 2 \\ \hline
         Spin  & $S$ & $3/2$\\ \hline
         Magnetic moment & $\smmu=g_{e}S\muB$      & 3 $\muB$  \\ \hline
         Isotropic bilinear exchange & \makecell{ $J_1$ \\ $J_2$ \\ $J_3$ }  & \makecell{\SI{1.0}{meV} \\ \SI{0.32002}{meV} \\ \SI{0.0081}{meV}} \\ \hline
         Anisotropic bilinear exchange & \makecell{$\lambda_1$  \\ $\lambda_2$\\ $\lambda_3$}     & \makecell{ \SI{0.1068}{meV} \\\SI{-0.01024}{meV} \\ \SI{0.0091}{meV}} \\ \hline
         Biquadratic exchange & $K_{bq}$        & \SI{0.21}{meV} \\ \hline
        Easy axis anisotropy & $D_z$ & \SI{0.10882}{meV} \\  \hline
        DM strength & $A$ & \SI{0.31}{meV} \\
        %Kitaev strength & $K$ & \SI{1}{meV} \textcolor{red}{good question!}
        \end{tabular}
        \end{ruledtabular}
    \end{minipage}
\end{table}

\section{Compatibility of the numerical results in the DM model with analytical solution and scan of DM angle direction \label{DMIappendix}}

\begin{figure}[h]
        \hspace{1.8cm}\begin{overpic}[scale=0.24]{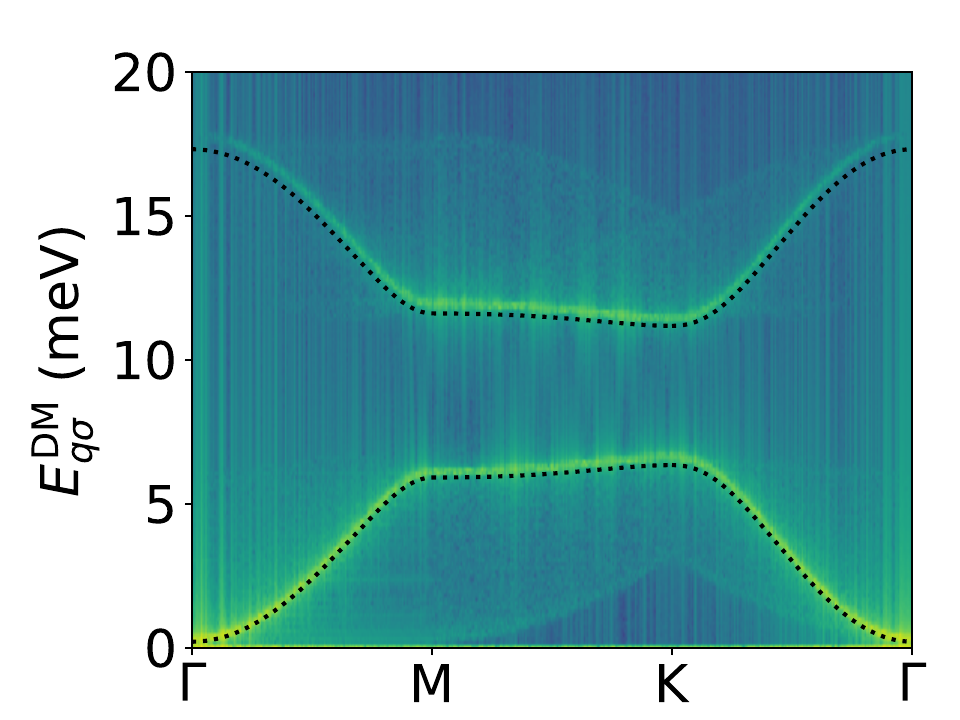}
        \put(-50,35) {\color{black} {\large $\theta_\text{DM}=0^\circ$}}
        \put(40,70) {\color{black} {\large first NN $J$}}
        \put(60,85) {\color{black} {\large OOP magnetization}}
        \end{overpic}
        \begin{overpic}[scale=0.24]{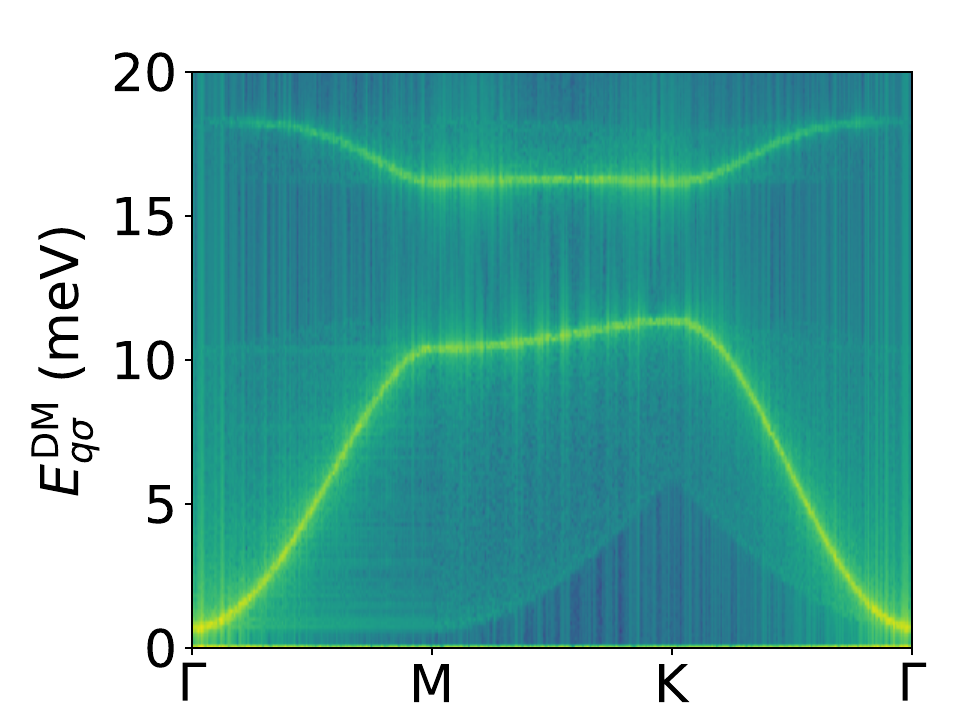}
        \put(40,70) {\color{black} {\large all NN $J$}}
        \end{overpic}
        \begin{overpic}[scale=0.24]{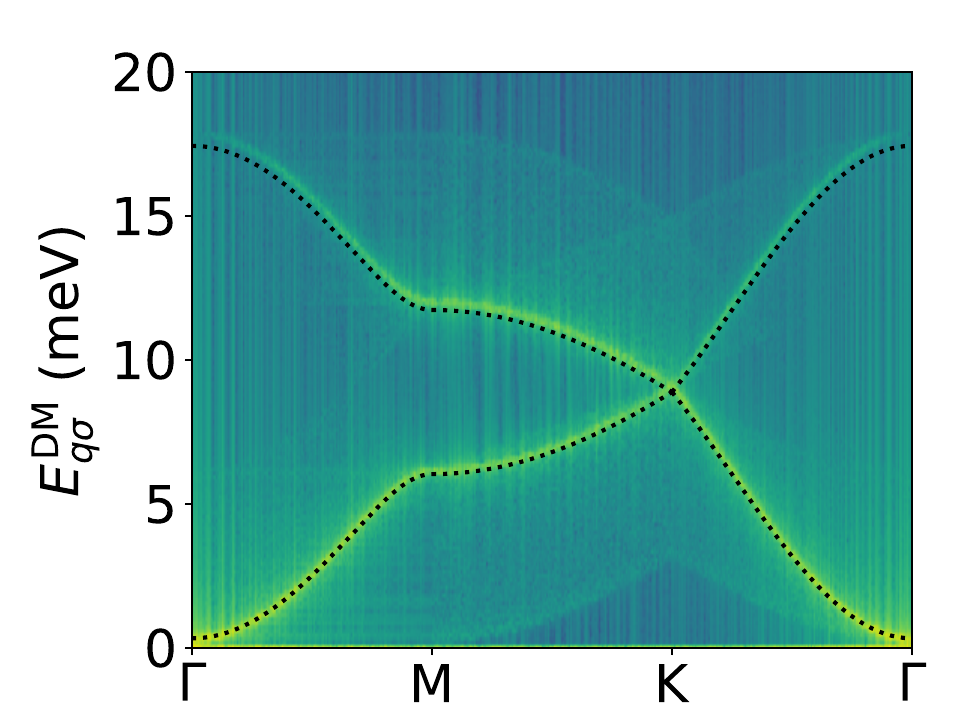}
        \put(40,70) {\color{black} {\large first NN $J$}}
        \put(65,85) {\color{black} {\large IP magnetization}}
        \end{overpic}
        \begin{overpic}[scale=0.24]{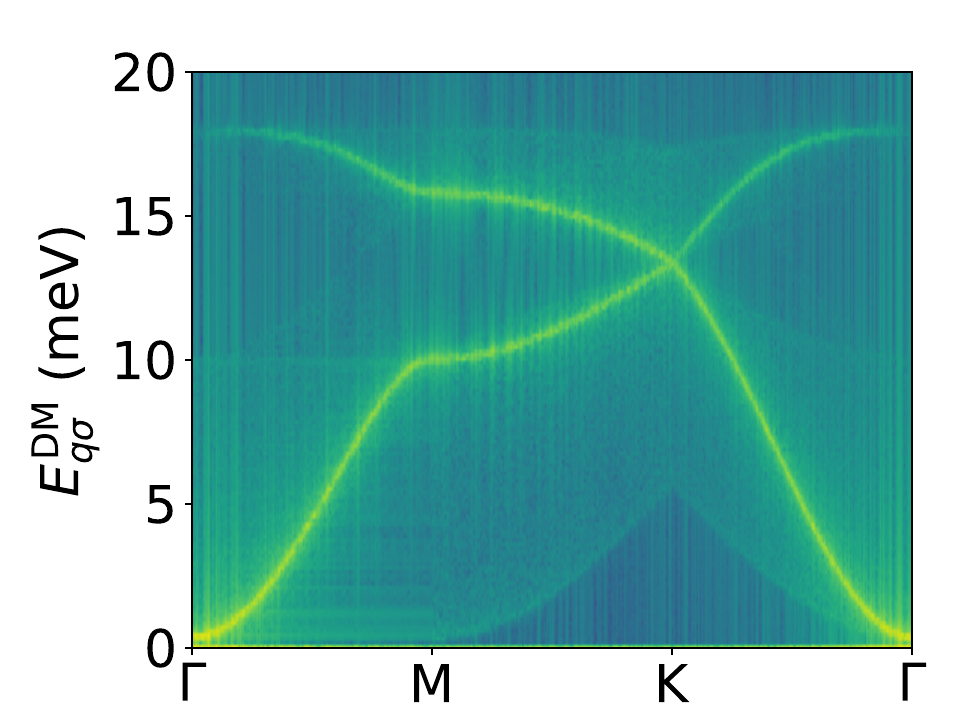}
            \put(40,70) {\color{black} {\large all NN $J$}}
        \end{overpic}\\
        \hspace{1.8cm}\begin{overpic}[scale=0.24]{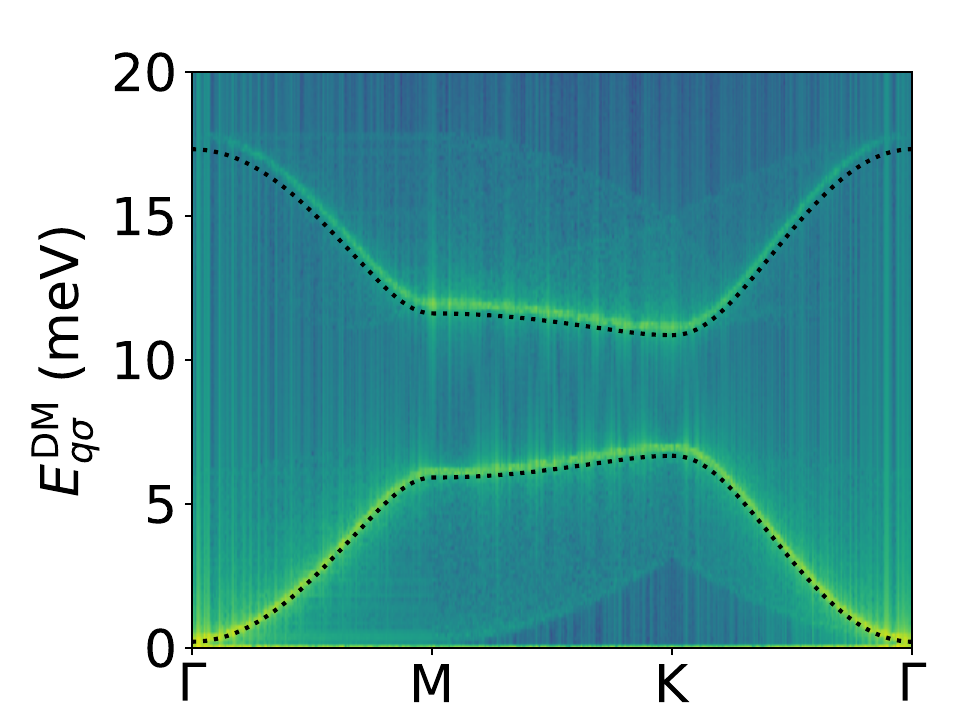}
        \put(-50,35) {\color{black} {\large $\theta_\text{DM}=30^\circ$}}
        \end{overpic}
        \includegraphics[width=0.22\linewidth]{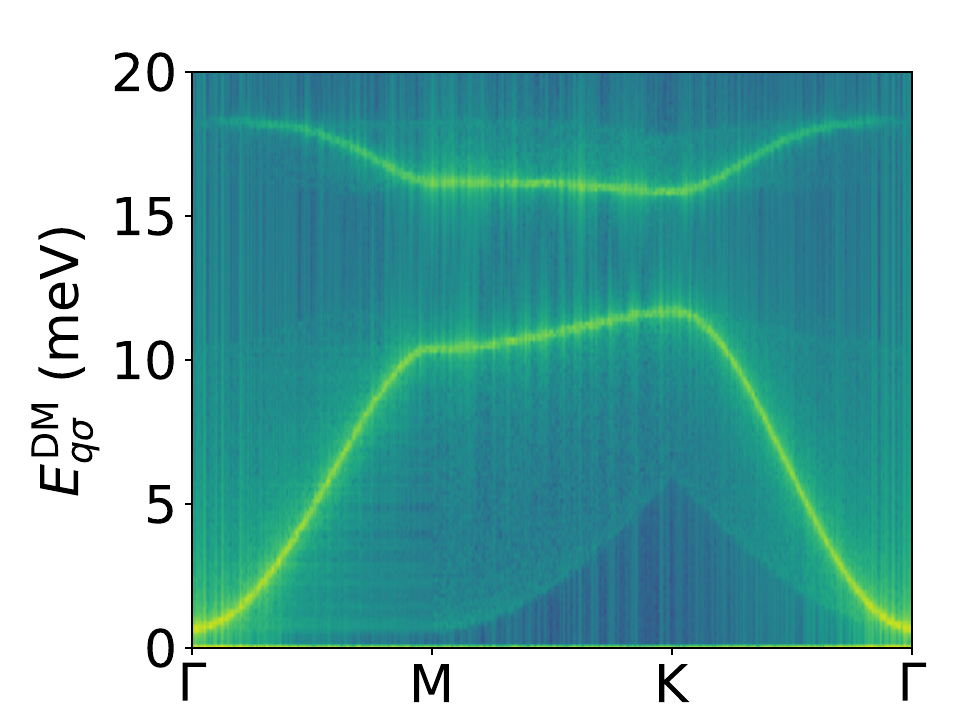}
        \includegraphics[width=0.22\linewidth]{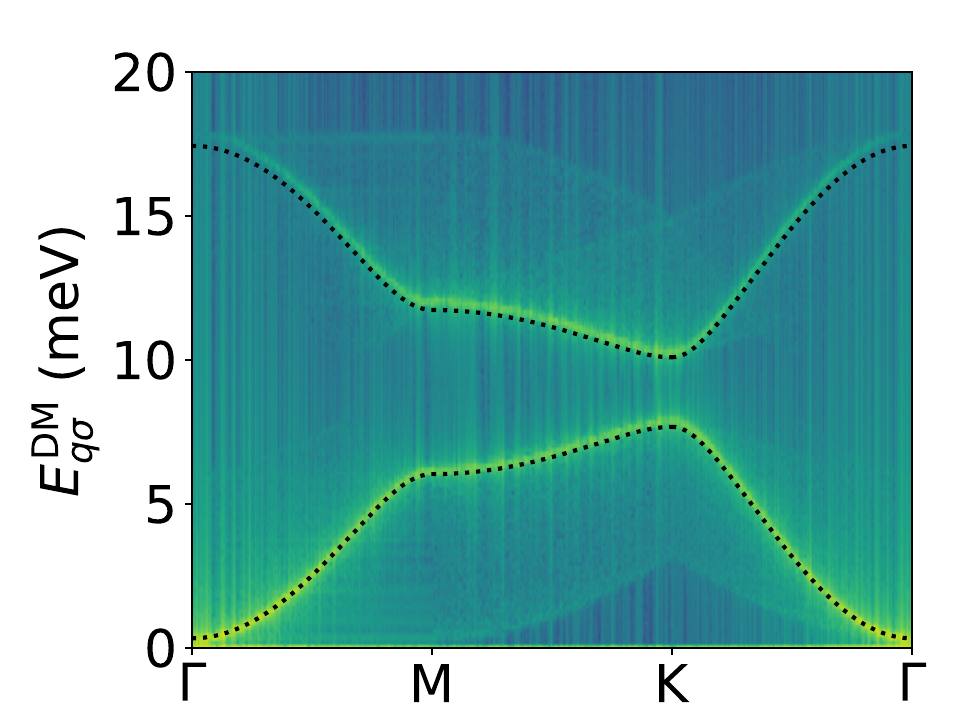}
        \includegraphics[width=0.22\linewidth]{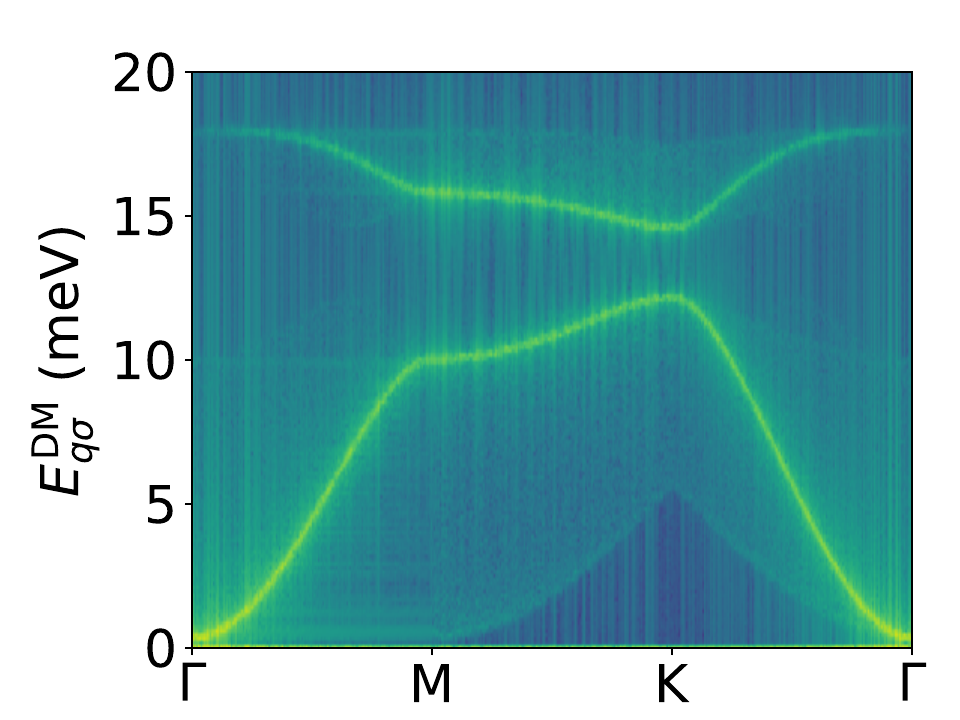}\\
        \hspace{1.8cm}\begin{overpic}[scale=0.24]{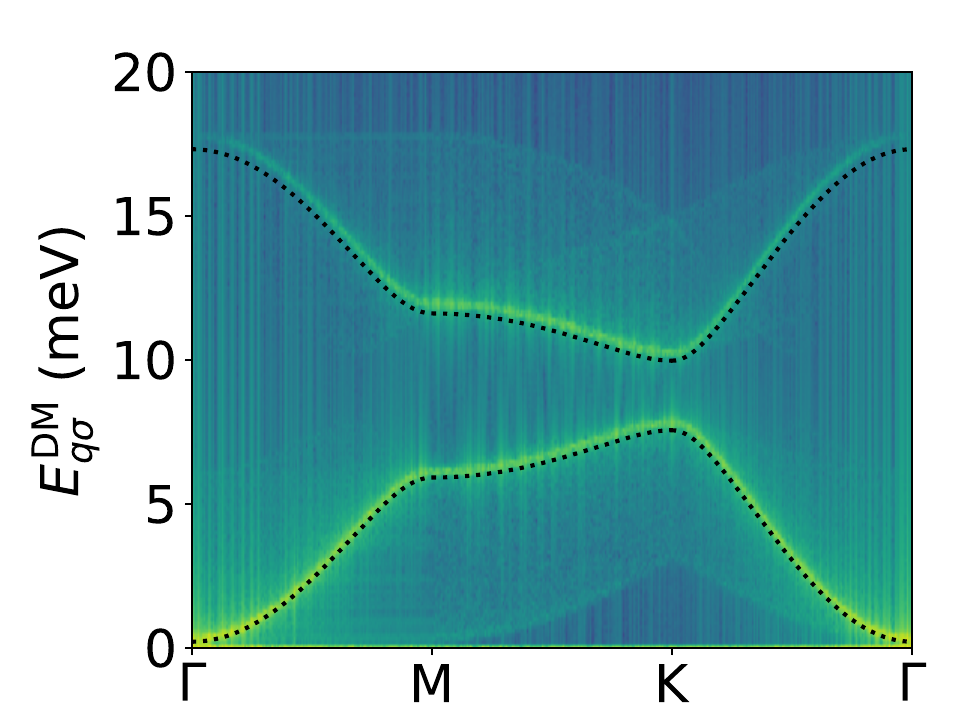}
        \put(-50,35) {\color{black} {\large $\theta_\text{DM}=60^\circ$}}
        \end{overpic}
        \includegraphics[width=0.22\linewidth]{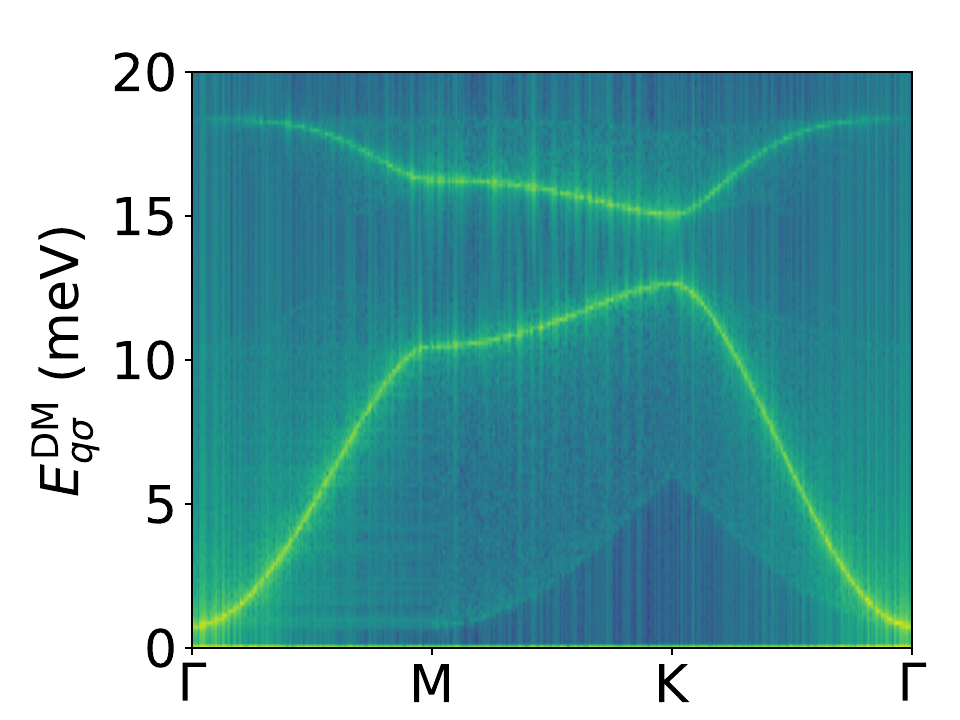}
        \includegraphics[width=0.22\linewidth]{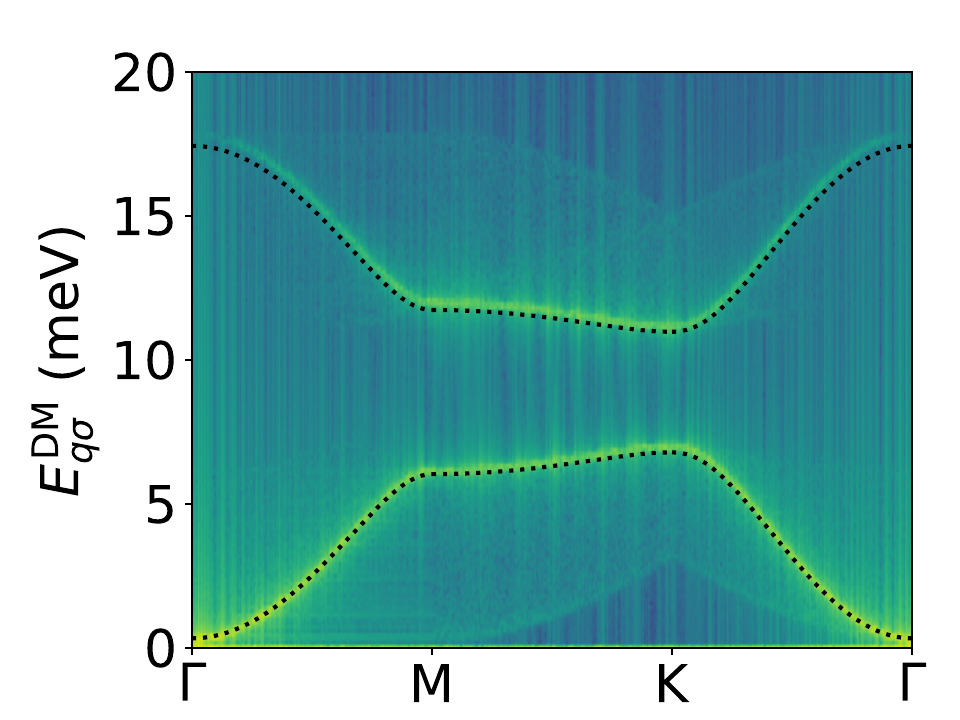}
        \includegraphics[width=0.22\linewidth]{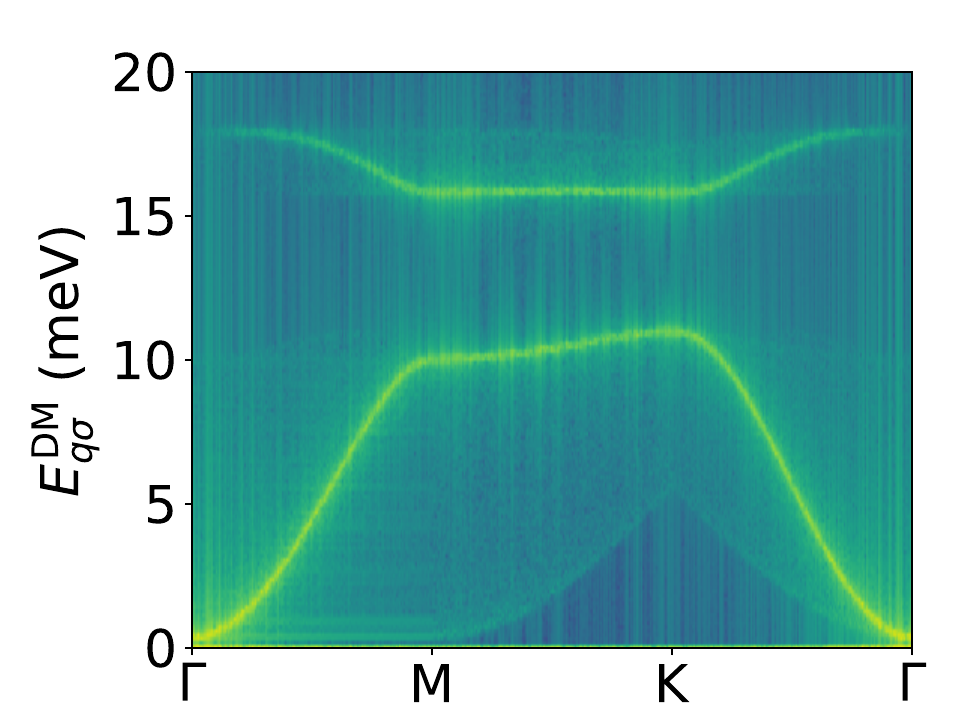}\\
        \hspace{1.8cm} \begin{overpic}[scale=0.24]{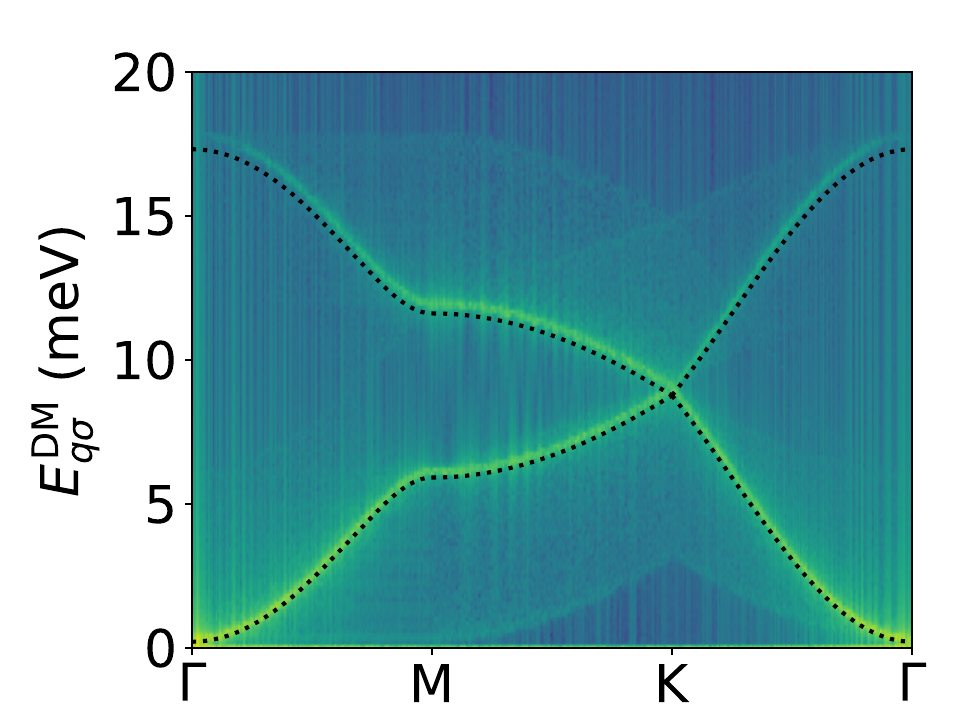}
        \put(-50,35) {\color{black} {\large $\theta_\text{DM}=90^\circ$}}
        \end{overpic}
        \includegraphics[width=0.22\linewidth]{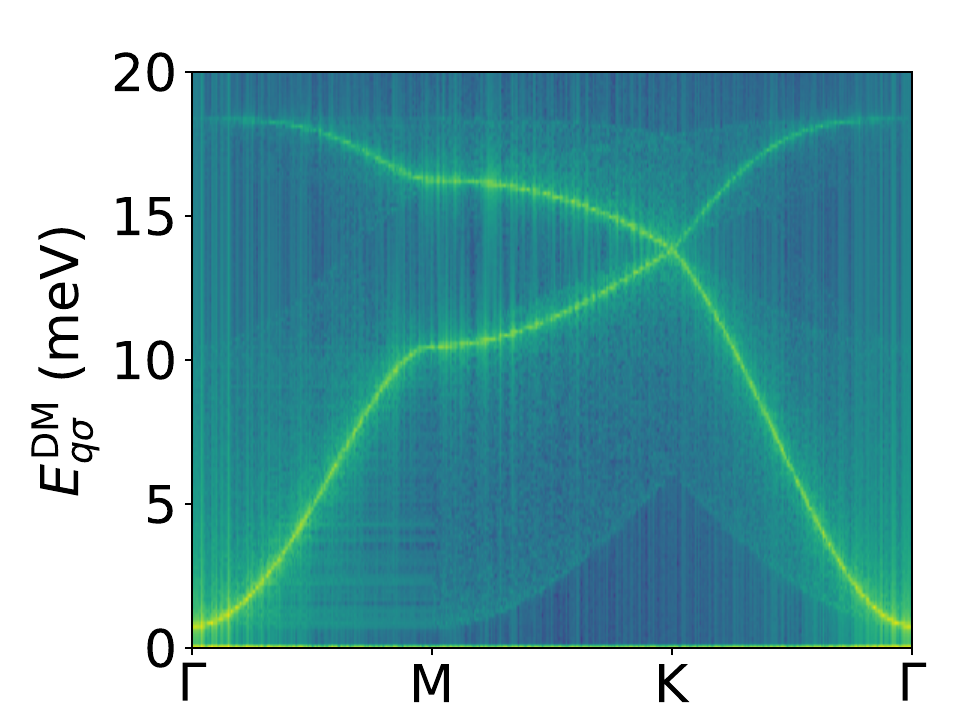}
        \includegraphics[width=0.22\linewidth]{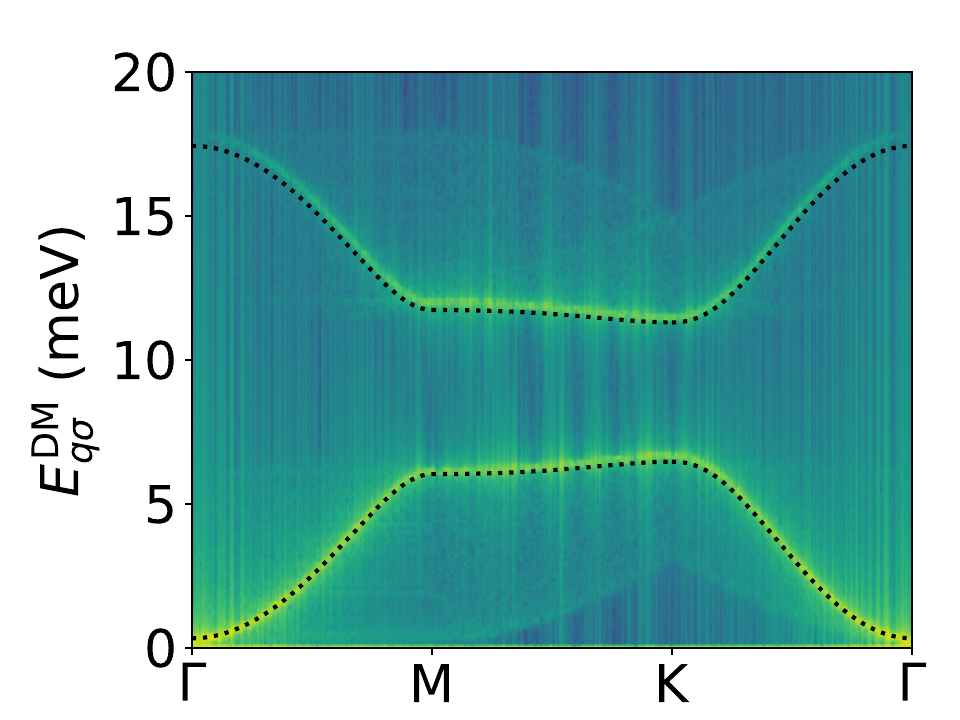}
        \includegraphics[width=0.22\linewidth]{graphics/DM_IP_AllNN_theta90.pdf} 
    \caption{Overview of magnon dispersion in the DM model at different DM angles $\theta_\text{DM}$ (increasing from top to bottom row) with OOP magnetization (left side) and IP magnetization (right side). For each magnetization configuration, numerical data is compared  when including only first-nearest neighbor (NN) exchange $J$ in the respective left columns to including all neighbors in the respective right columns. Analytical data is only found for first-nearest neighbor and shown with black dashed lines.}
    \label{fig:fullOverviewDM}
\end{figure}

The purpose of this appendix is to demonstrate the compatibility of the analytical solution with the numerical result in the DM model, and to show a full scan of the dispersion relation in the IP and OOP configuration when tilting the DM vector. A full overview of the magnon dispersion dependent on the DM direction can be found in \cref{fig:fullOverviewDM}. Both with OOP and IP magnetization (left to columns compared to right two columns), the gap closes when the DM vector is orthogonal to the magnetization direction (that is, $\theta_\text{DM}=90^\circ$ at OOP magnetization and $\theta_\text{DM}=0^\circ$ at IP magnetization). The numerical results are shown with color maps and compared to the analytical results (dashed black lines) for a model where only first nearest-neighbor (NN) exchange $J$ is included (respective left column for each magnetization direction). Here, analytical and numerical results show perfect agreement. Including more nearest neighbor interactions (respective right column for each magnetization direction) leads to a flattening of the high-energy branch and stretching of the low-energy branch, but has no impact on the gap size. Note that tilting the DM direction is equivalent to tilting the magnetization direction and fixing the DM vector direction, since only the relative angle between magnetization and DM vector is relevant for the gap.

\vspace{1cm}

\section{Kitaev model under variation of an external magnetic field angle \label{KitaevAppendix}}
\begin{figure}[h]
    \centering
    \includegraphics[trim = 0 0 50 0,width=0.16\linewidth]{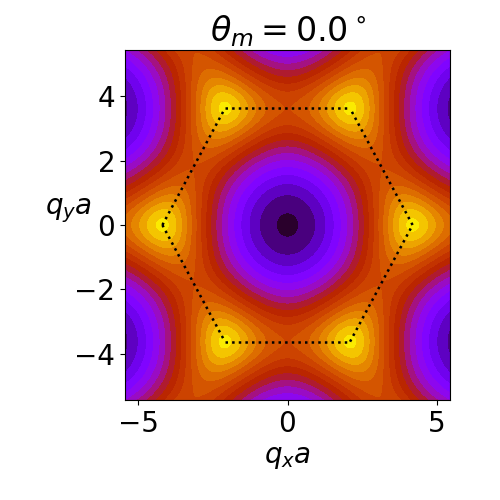}
    \includegraphics[trim = 0 0 50 0,width=0.16\linewidth]{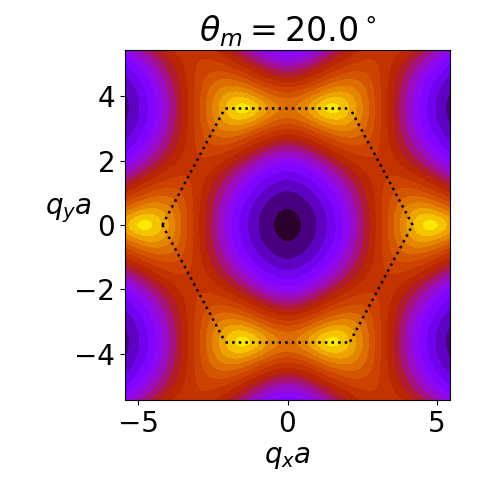}
    \includegraphics[trim = 0 0 50 0,width=0.16\linewidth]{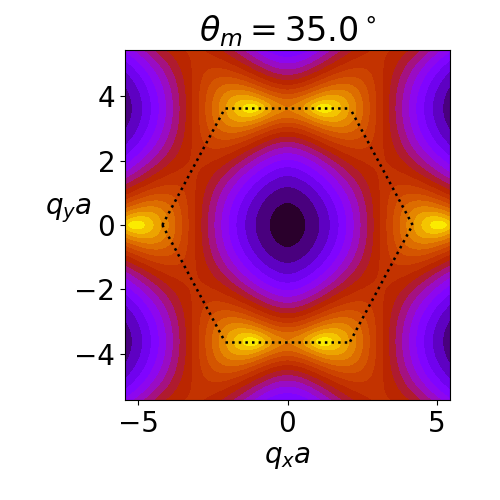}
    \includegraphics[trim = 0 0 50 0,width=0.16\linewidth]{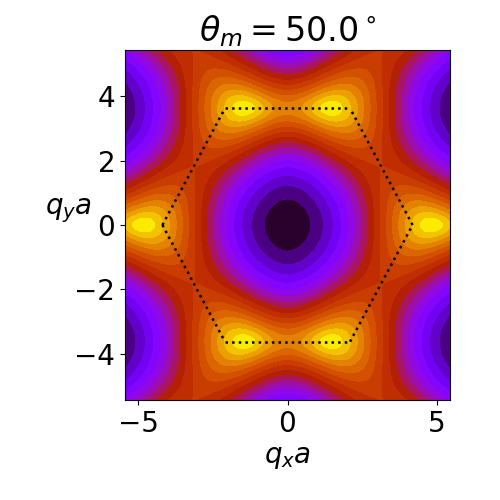}
    \includegraphics[trim = 0 0 50 0,width=0.16\linewidth]{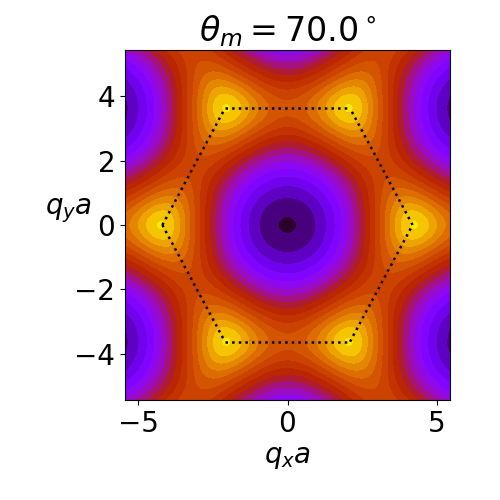}
    \includegraphics[trim = 0 0 50 0,width=0.16\linewidth]{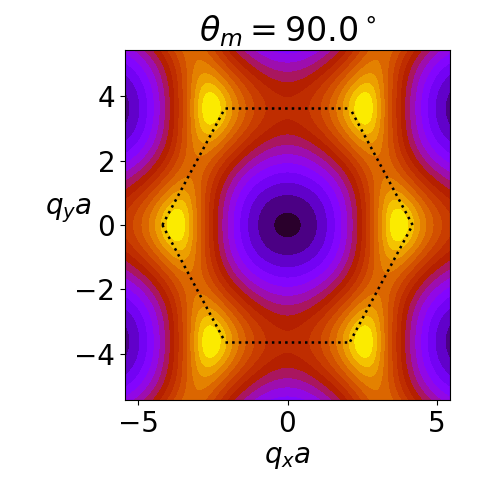}
    \caption{Migration of the maxima (yellow) of the lower branch of the dispersion relation in the Kitaev model. Black dashed lines indicate the Brillouin zone.}
    \label{fig:KitaevContoursLowerBranch}
\end{figure}

In this section, the magnetization direction dependency of the dispersion relation in the Kitaev model is presented. The migration of the Dirac points with increasing magnetization angle $\theta_m$ from left to right is shown in \cref{fig:KitaevContoursLowerBranch}. The color map refers to the energy, with yellow being the maximum. The dashed black lines indicate the Brillouin zone. While in the OOP configuration, $\theta_m=0^\circ$, the maxima are at the K points, they migrate towards the M points up til $\theta_m=35^\circ$, then back towards K at $\theta_m=70^\circ$ and then out of the first Brillouin zone. This shift of the Dirac point is shown in \cref{fig:KitaevGapClosing}(a) in the lower panel with the green dashed line ($\sigma=-$ denotes the lower branch). Further details are provided in the main text.

As in the DM model, the Dirac gap is closed at certain magnetization directions. The full dispersion relation per magnetic field direction is presented in \cref{fig:KitaevGapClosing}(b), where the color coding represents the magnetization angle from OOP (blue) to IP (yellow). At $\theta_m=35^\circ$, the gap is closed, but not at the K point, due to the shift of Dirac points. In \cref{fig:KitaevGapClosing}(a) in the upper panel, the dependency of the minimal gap on the magnetization direction is shown with the purple dotted line. The numerically found dispersion relation at $\theta_m=35^\circ$ is shown in \cref{fig:KitaevGapClosing}(c).

\begin{figure}
    \centering
    \begin{overpic}[scale=0.4]{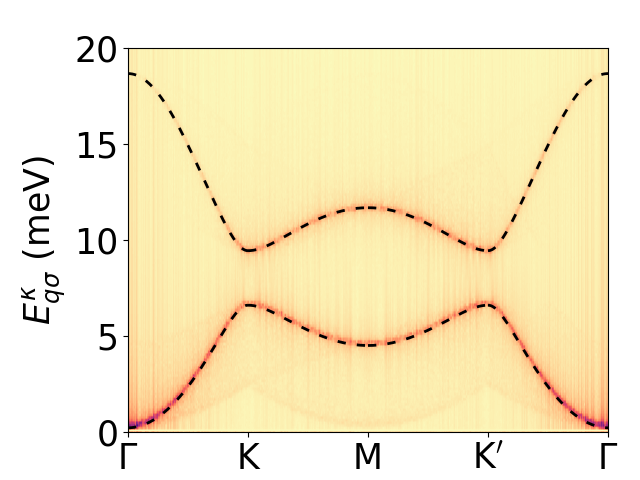}
        \put(1,70) {\color{black} (a)}
    \end{overpic}
    \begin{overpic}[scale=0.4]{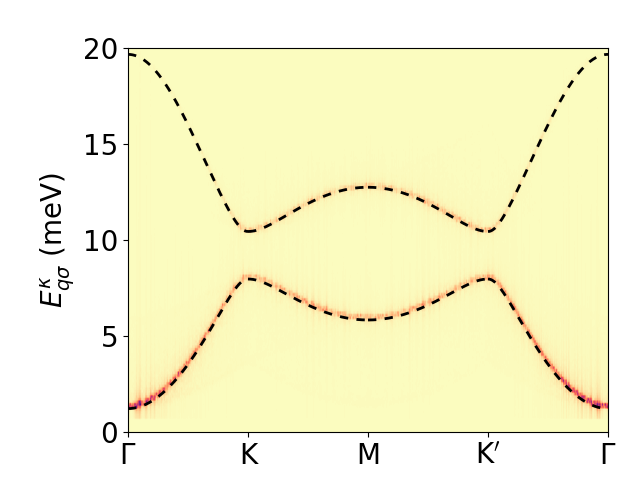}
        \put(1,70) {\color{black} (b)}
    \end{overpic}
    \caption{Magnon dispersion relation in the OOP configuration in the Kitaev model (a) without applied magnetic field, and (b) with applied magnetic field along the OOP direction with strength \SI{9}{T}. We read out a reduction of the gap by around 13\% from $\Delta_\text{K}(0)=\SI{2.7}{meV}$ to $\Delta_\text{K}(0)=\SI{2.35}{meV}$.}
    \label{fig:KitaevMagneticField}
\end{figure}

\begin{figure}[h]
    \centering
        \begin{overpic}[trim=0 20 0 0 , scale=0.31]{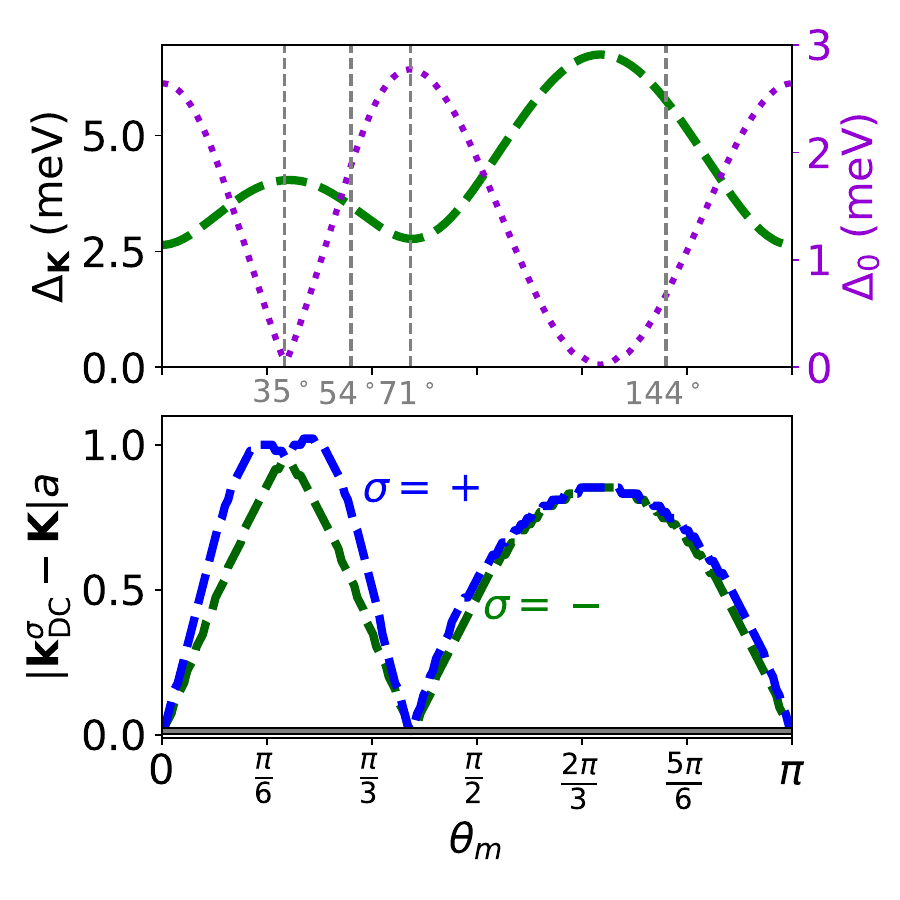}
        \put(1,89) {\color{black} (a)}
        \end{overpic}
        \begin{overpic}[scale=0.4]{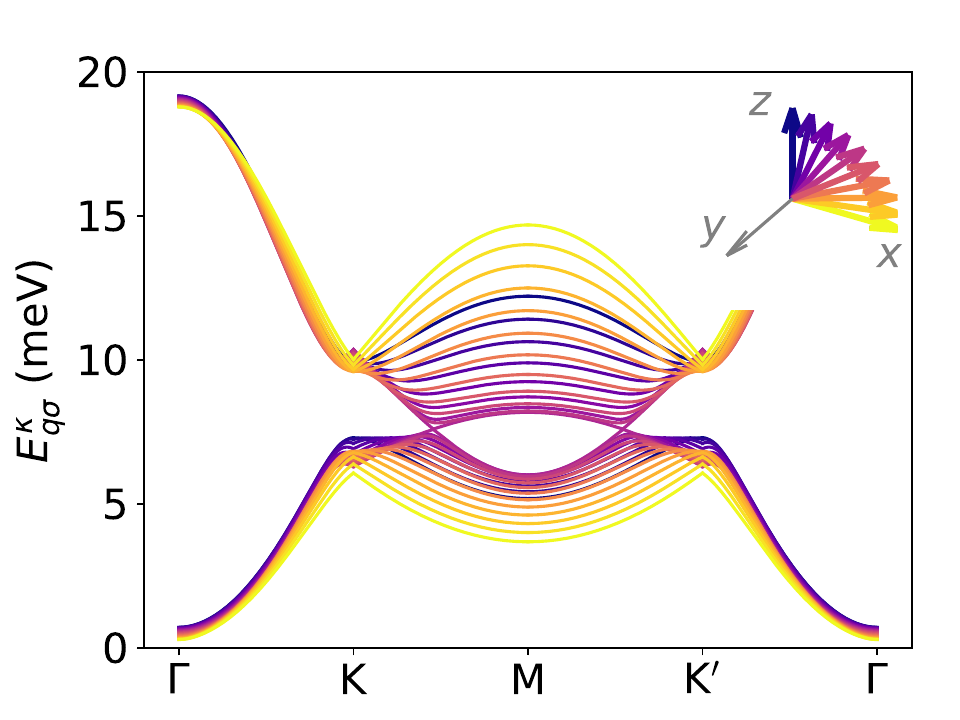}
        \put(-1,65) {\color{black} (b)}
        \end{overpic}
        \begin{overpic}[scale=0.4]{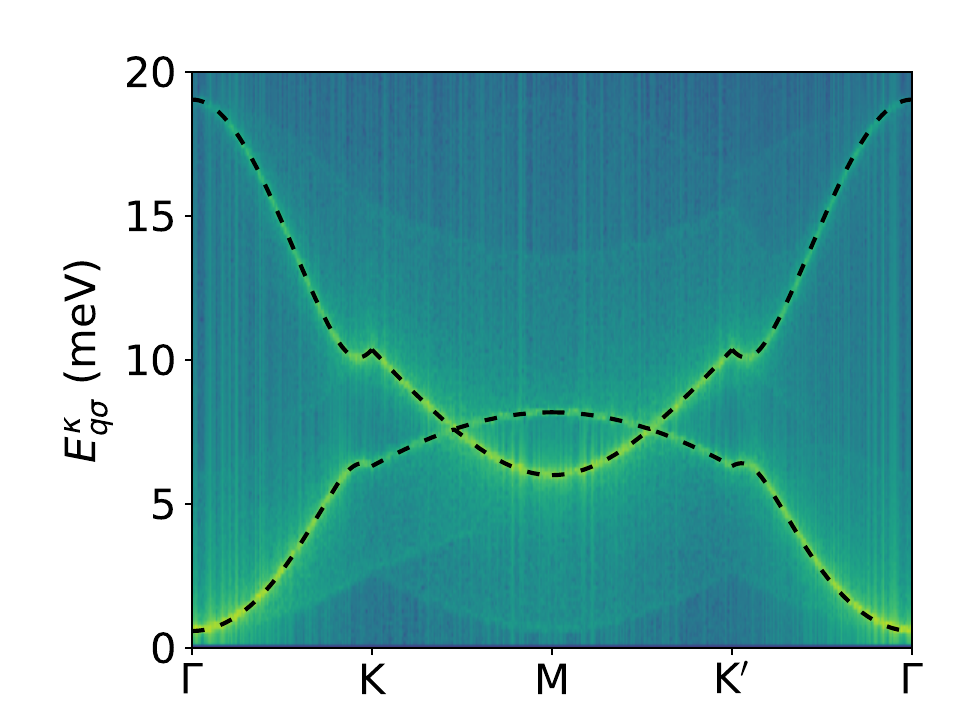}
        \put(5,65) {\color{black} (c)}
        \end{overpic}
    \caption{Impact of tilting angle $\theta_m$ of the magnetization on the magnon dispersion in the Kitaev model. (a) Upper panel: gap at the K point $\Delta_K$ (green dashed line, referring to the left axis) and minimal gap $\Delta_0$ (purple dotted line, referring to the right axis). This minimal gap is shifted away from the K point as demonstrated in the lower panel. There, the displacement of the Dirac points in the acoustic and optical branch from the K-point $(|\mathbf{k}_{\rm{DC}} - \rm{\bm{K}}|a)$ is shown. In our Kitaev model, the gap closes at $\theta_m=35^\circ$. (b) Full dispersion relations depending on the magnetization direction which is tilted from OOP (blue) to IP (yellow). (c) Numerically found dispersion relation (color map) along the analytical solution (black dashed line) at $\theta_m=35^\circ$}
    \label{fig:KitaevGapClosing}
\end{figure}

Furthermore, we show a numerical verification of the impact of the magnetic field strength on the magnon dispersion relation in the Kitaev model. In \cref{fig:KitaevMagneticField}(b), a magnetic field with strength \SI{9}{T} is applied, reducing the K gap compared to the ground state without magnetic field as shown in \cref{fig:KitaevMagneticField}(a).

\section{Analytical calculations}\label{A:analyticalCalc}
In this section, we derive the magnon dispersion relation for both the Kitaev and DM model using linear spin wave theory at zero temperature.

\subsection{DM model}
In our analytical calculations, we reduce Eq. (2) from the main text to

\begin{align}
\label{eqn: magnon_general_spin_in_plane_Hamiltonian}
    \ham = &- J \sum_{<i,j>} \sms_i \cdot \sms_j - K_\text{bq} \sum_{<i,j>} (\sms_i \cdot \sms_j)^2 - \smmu h_0 \sum_i \smB \cdot \sms_i  - D_z\sum_i S_{iz}^2 - \sum_{i,j} \smA_{ij} \cdot [\gv{S}_i \times \gv{S}_j].
\end{align} %translate Kartsev to Jostein: \Lambda = K_bq, mu_s h B = h, D = K_z, A = D_ij

where only isotropic first nearest neighbor interaction $J$ is taken into account. All other constants are defined in the main text.  
Since the unit cell carries two atoms, the lattice has to be split in two sublattices \texttt{A} and \texttt{B}.
The DM interaction we express as
\begin{align}
        &\smA^{\text{NNN}}_{ij} = \nu_{ij} (A_x^{\text{NNN}} \gv{\hat{x}} + A_y^{\text{NNN}} \gv{\hat{y}} + A_z^{\text{NNN}} \gv{\hat{z}})
\end{align}
where $\nu_{ij}$ is the Haldane sign. 

The Holstein-Primakoff transformation can be expressed in linear order through the bosonic creation and annihilation operators $\hat{a}$ and $\hat{a}^\dagger$ as
\begin{equation}
    \begin{split}
        S_{i\texttt{A}1} =& \sqrt{\frac{S}{2}} (\hat{a}_i + \hat{a}_i^\dagger), \\
        S_{i\texttt{A}2} =& -\iu \sqrt{\frac{S}{2}} (\hat{a}_i - \hat{a}_i^\dagger), \\
        S_{i\texttt{A}3} =& S - \hat{a}_i^\dagger \hat{a}_i,
    \end{split}
\end{equation}
and similarly for the spins at sublattice $\texttt{B}$. 

In order to find the dispersion for a general magnetization direction, we switch basis to a rotated frame of reference $\hat{e}_1 = O \hat{x}, \hat{e}_2 = O \hat{y}, \hat{e}_3 = O \hat{z}$ through a rotational matrix $O$ with 
\begin{align*}
    O_z(\theta) &= \begin{bmatrix}
        \cos(\theta) & -\sin(\theta) & 0 \\
        \sin(\theta) &\cos(\theta) & 0 \\
        0 & 0 & 1
    \end{bmatrix}, 
    O_y(\phi) = \begin{bmatrix}
        \cos(\phi) & 0 & \sin(\phi)\\
        0 & 1 &0 \\
        -\sin(\phi) & 0 &\cos(\phi)
    \end{bmatrix}, 
    O_x(\theta) = \begin{bmatrix}
        1 & 0 & 0 \\
        0 & \cos(\theta) & -\sin(\theta)\\
        0 & \sin(\theta) & \cos(\theta)
    \end{bmatrix} \nonumber \\
     O &= O_z(\theta)O_y(\phi)O_x(\theta)= \begin{bmatrix}
        O_{x1} & O_{x2} & O_{x3} \\
        O_{y1} & O_{y2} & O_{y3} \\
        O_{z1} & O_{z2} & O_{z3} \\
    \end{bmatrix} 
\end{align*}
The rotation matrix is an orthogonal matrix satisfying $O^T O=I$. The new frame of reference is defined such that $\hat{e}_3$ is aligned with the magnetic field, and hence the magnetization direction. We assume the magnetic field strength is large enough to set the ground state.
With the structure factor 
\begin{equation}
    |f_{\delta_A}(\gv{k})|^2 = 1 + 4\cos\left(\frac{1}{2}k_x a\right)\left[\cos\left(\frac{1}{2}k_x a\right) + \cos\left(\frac{\sqrt{3}}{2}k_y a\right) \right]
\end{equation}
for $z=3$ nearest neighbors and the lattice constant $a$, and defining the abbreviations
\begin{align}
    \Delta' &= \smmu h_0 + \left((2S-1) O_{z3}^2 - S(O_{z1}^2 + O_{z2}^2)\right) \Tilde{D_z}  \label{A:EqAbbreviations1}\\
    \Tilde{J} &= J + 2K_\text{bq} S (S-1) \label{A:EqAbbreviations2} \\ 
    \upsilon &= (\Tilde{J} Sz + \Delta')^2  \label{A:EqAbbreviations3}\\
    \Tilde{D_z} &= S^2 D_z^2 |Q_z|^2 \label{A:EqAbbreviations4} \\
    Q_z &= (O_{z1} - \iu O_{z2})^2, \label{A:EqAbbreviations5}
\end{align}
we can find the eigen energies with the rotated DM vector 
\begin{equation}
    \begin{split}
        \smA_k^{\text{NNN}} =& -2 (O_{x3} A_x^{\text{NNN}} + O_{y3} A_y^{\text{NNN}} + O_{z3} A_z^{\text{NNN}}) \left[\sin(k_x a) - 2 \sin\left(\frac{1}{2}k_x a\right) \cos\left(\frac{\sqrt{3}}{2} k_y a\right)\right]
    \end{split}
\end{equation}
as
\begin{align}
        E^{\pm}_k = &\left( \upsilon + S^2 (\smA_k^{\text{NNN}})^2 + \Tilde{J}^2 S^2|f_{\delta_A}(\gv{k})|^2 - \Tilde{D_z} \pm 2S \sqrt{(\smA_k^{\text{NNN}})^2 (\upsilon - S^2 D_z^2 |Q_z|^2) + \Tilde{J}^2 |f_{\delta_A}(\gv{k})|^2A} \right)^{1/2}. 
\end{align}

In the OOP configuration and OOP DM  direction ($\theta=0$), the gap at the K point $\left(2 \pi/(3a) ,2\pi/\sqrt{3a}\right)$ can be found from $E^\pm(K) = JS\left(z \pm \sqrt{\frac{A^2(K)}{\Tilde{J}^2}} - f(K) \right) + \Delta $ with the effective coupling $\Tilde{J}$ that contains the biquadratic coupling, the gap $\Delta = \mu h_0 + D_z$, the DM strength $A$ and the number of NN $z$. At the K point, the structure factor is 0, which means the gap expression reduces to $E^+(K) - E^-(K) = \Tilde{J}S\left(z + \frac{A(K)}{\Tilde{J}}\right) - \Tilde{J}S\left(z - \frac{A(K)}{\Tilde{J}}\right) = 2SA(K)$. Here, $A(K)= 2A(\sin(K_x a) - 2 \sin(\frac{1}{2}K_x a) \cos(\frac{\sqrt{3}}{2} K_y a)) = 3A\sqrt3$. 
With our parameters $A=\SI{0.31}{meV}$ and $S=1.5$, this leads to $\Delta_K = \SI{4.8}{meV}$, which we define as $\Delta_0$ in our fit in Fig. 3 in the main text. 

\subsection{Kitaev model}

For the Kitaev model, we write Eq. (1) from the main text as $\ham = \ham_\kappa + \ham_0$ where $\ham_0 = - D_z \sum_i \left(\sms_i \cdot \hat{z}\right)^2 - \smmu h_0 \sum_i \smB \cdot \sms_i $ and the interaction term $\ham_\kappa$ includes both nearest neighbor Heisenberg and Kitaev interaction. In the local Kitaev frame, the interaction term reads \cite{Geometrical_frustration_versus_Kitaev_interactions}
\begin{gather}
    \ham^\prime_\kappa = -\sum_{<i,j> \in \gamma } \Tilde{\sms}_i^T H_{\kappa,\gamma} \Tilde{\sms}_j \\
    H_{\kappa,\xi} = \begin{bmatrix}
        J+\kappa & 0 & 0 \\
        0 & J & 0 \\
        0 & 0 & J \\
    \end{bmatrix}, \nonumber  H_{\kappa,\eta} = \begin{bmatrix}
        J & 0 & 0 \\
        0 & J+\kappa & 0 \\
        0 & 0 & J \\
    \end{bmatrix}, \nonumber  H_{\kappa,\zeta} = \begin{bmatrix} 
        J & 0 & 0 \\
        0 & J & 0 \\
        0 & 0 & J+\kappa \\
    \end{bmatrix}, \nonumber 
\end{gather}
with $\kappa$ as the Kitaev interaction strength. The notation $\langle i,j\rangle \in \gamma$ symbolizes the nearest neighboring bonds between sublattice $A$ and $B$, and the spins $\Tilde{\sms}$ are written in the local Kitaev frame $\{\gv{\hat{\xi}}, \gv{\hat{\eta}}, \gv{\hat{\zeta}}\}$. The neighboring bonds are denoted by $\gamma \in \{\xi, \eta, \zeta\}$. In CrI$_3$, each Cr-Cr bond has an associated Kitaev axis directed towards the intermediate I$^-$ ion, based on the direction of the super-exchange path. The position of these Kitaev axes relative to the Cartesian (crystallographic) axes is given by \cite{Geometrical_frustration_versus_Kitaev_interactions}

\begin{equation*}
    \gv{\hat{\xi}} = \begin{pmatrix}\frac{1}{\sqrt{6}}\\-\frac{1}{\sqrt{2}}\\\frac{1}{\sqrt{3}}\end{pmatrix}, \hspace{2mm} \gv{\hat{\eta}} = \begin{pmatrix}\frac{1}{\sqrt{6}}\\ \frac{1}{\sqrt{2}} \\ \frac{1}{\sqrt{3}}\end{pmatrix}, \hspace{2mm} \gv{\hat{\zeta}} = \begin{pmatrix}-\frac{\sqrt{6}}{3} \\ 0 \\\frac{1}{\sqrt{3}}\end{pmatrix},
\end{equation*}
suggesting that the spins in the Cartesian frame are related to the spins in the Kitaev frame via $\gv{\Tilde{S}}_i = U_\kappa \gv{S}_i$, where $U_\kappa$ is an orthogonal matrix containing $\gv{\hat{\xi}}$, $\gv{\hat{\eta}}$ and $\gv{\hat{\zeta}}$ as rows. By performing this transformation, we obtain 
\begin{equation*}
    \ham_K = - \sum_{<i,j> \in \gamma} \sms_i^T H_{C,\gamma} \sms_j,
\end{equation*}
with $H_{C,\gamma} = U_\kappa^T H_{\kappa,\gamma} U_\kappa$. Computing the matrices for each bond, the summation may be written compactly as \cite{ChenKee2023_Kitaev}

\begin{gather}
    \ham_K = - J' \sum_{<i,j> \in \gamma} \sms_i \cdot \sms_j  + \frac{1}{3}\kappa \sum_{<i,j> \in \gamma} \sms_i^T W^{\gamma} \sms_j, \nonumber 
    W^{\gamma} = \begin{bmatrix}
        -c_{\phi_{\gamma}} & s_{\phi_{\gamma}} & \sqrt{2} c_{\phi_{\gamma}} \\
        s_{\phi_{\gamma}} & c_{\phi_{\gamma}} & \sqrt{2} s_{\phi_{\gamma}} \\
        \sqrt{2} c_{\phi_{\gamma}} & \sqrt{2} s_{\phi_{\gamma}} & 0 \\
    \end{bmatrix}, \nonumber
\end{gather}
where $J' = J_\kappa + \frac{1}{3}\kappa$, $c_{\phi_{\gamma}} = \cos(\phi_{\gamma})$, $s_{\phi_{\gamma}} = \sin(\phi_{\gamma})$ and $\phi_{\gamma} = 0, 2\pi/3, 4\pi/3$ for $\gamma = \zeta, \xi, \eta$ bonds respectively. The first term represents a scaling of the isotropic Heisenberg exchange coupling coefficient, while the second term introduces anisotropic contributions. As before, the spins are rotated via $\sms_{i} = O \sms'_i$ to align with the external magnetic field. The matrix multiplication in the second term can then be expressed as $\sms'^T_i M^{\gamma} \sms'_j$, where $M^{\gamma} = O^T W^{\gamma} O$. After using a Holstein-Primakoff transformation and retaining only second-order terms, we insert Fourier-transformed magnon operators and add an on-site anisotropy and Zeeman term as in the DM model. Finally, the positive eigenvalues are found with a Bogoliubov-de-Gennes as

\begin{equation}
    E_k^{\pm} = \frac{1}{\sqrt{2}} \left( -\Tilde{A}_k \pm \sqrt{\Tilde{A}_k^2 - 4 \Tilde{B}_k} \right)^{\frac{1}{2}},
\end{equation}
with the helping variables defined as

\begin{align}
    \Tilde{A}_k =& -2t_0^2 - 2|\Tilde{t}_k|^2 + 2 S^2 D_z^2 |Q_z|^2 + |d_k|^2 + |d_{-k}|^2, \label{eq:atilde}\\
    \Tilde{B}_k =& (t_0^2 - |\Tilde{t}_k|^2)^2 + S^2 D_z^2 |Q_z|^2 (S^2 D_z^2 |Q_z|^2 - 2 t_0^2 - 2 |\Tilde{t}_k|^2) - t_0^2 (|d_k|^2 + |d_{-k}|^2) + |d_k|^2 |d_{-k}|^2 \nonumber\\
    & + 4 t_0 S D_z \real{\Tilde{t}_k Q_z d_{-k}^* + \Tilde{t}_k Q_z^* d_k} - 2 \real{\Tilde{t}_k^2 d_k d_{-k}^*} - 2S^2 D_z^2 \real{Q_z^2 d_k^* d_{-k}^*} \label{eq:btilde}
\end{align}

using 

\begin{align}
    t_0 =& 3 J' S + \Delta', \label{eq:t0} \\
    \Tilde{t}_k =& J' S f_{\delta_A} (\gv{k}) - \frac{\kappa S}{6} \left[(\Gamma_{yy} - \Gamma_{xx} + 2 \sqrt{2} \Gamma_{xz}) \e^{\iu \frac{k_y a}{\sqrt{3}}} - \e^{-\iu \frac{k_y a}{2\sqrt{3}}} \left( (\Gamma_{yy} - \Gamma_{xx} + 2 \sqrt{2} \Gamma_{xz}) \cos\left(\frac{1}{2} k_x a\right) \right. \right. \nonumber \\ 
    &\left.\left. + \iu 2 \sqrt{3} (\Gamma_{xy} + \sqrt{2} \Gamma_{yz}) \sin\left(\frac{1}{2} k_x a\right) \right)\right], \label{eq:ttilde} \\
    d_k =& \frac{\kappa S}{6} \left[(O_y^2 - O_x^2 + 2\sqrt{2} O_x O_z) \e^{-\iu \frac{k_y a}{\sqrt{3}}} - \e^{\iu \frac{k_y a}{2\sqrt{3}}} \left( (O_y^2 - O_x^2 + 2\sqrt{2} O_x O_z) \cos\left(\frac{1}{2} k_x a\right) \right. \right. \nonumber \\
    &\left.\left.  - \iu 2 \sqrt{3} (O_x O_y + \sqrt{2} O_y O_z) \sin\left(\frac{1}{2} k_x a\right) \right)\right], \label{eq:dk}
\end{align}

where we defined the quantities $\Gamma_{\mu\nu} = O_{\mu 1} O_{\nu 1} + O_{\mu 2}O_{\nu 2}$ and $O_{\mu} = O_{\mu 1}- \iu O_{\mu 2}$ for shorthand notation.

In order to find an analytical expression of the Dirac gap in the OOP configuration, we set $\theta=0=\phi$, which leads to $Q_z=1$, $O_x=1,O_y=-i,O_z=0$ and $\Gamma_{ij}=\delta_{ij}$. The structure factor $f_{\delta_A}=0$ at the K point. Thus, \Cref{eq:t0,eq:ttilde,eq:dk} are reduced to
\begin{align*}
    t_0 &= 3J'S+\mu_s h_0 + 2D_z \\
    \Tilde{t} &= 0 \\
    d_k %&= \frac{KS}{6}\left[((-i)²-1) \exp(-iK_ya/\sqrt{3})-\exp(i K_ya/2\sqrt{3} \left(((-i)²-1)\cos(k_xa/2)-i2\sqrt{3}(-i)\sin(k_xa/2)\right) \right]\\
    %&= \frac{KS}{6}\left[-2\exp(-iK_ya/\sqrt{3})-\exp(i K_ya/2\sqrt{3} \left(-2\cos(k_xa/2)-2\sqrt{3}\sin(k_xa/2)\right) \right]\\
    &=  \frac{\kappa S}{6}\left[-2\exp(-i2\pi/3)-\exp(i \pi/3) \left(-2\cos(\pi/3)+2\sqrt{3}\sin(\pi/3)\right) \right]
\end{align*}
where the last leads to $|d_k|^2=(S \kappa)^2, |d_{-k}|^2 = 0$ and $d_kd_{-k}^\star=0$. With that, \Cref{eq:atilde,eq:btilde} read
    $\Tilde{A} = -2t_0^2 + 2S^2 D_z^2 +  (S \kappa)^2$ and 
    $\Tilde{B} = t_0^4 + S^4D_z^4 -2t_0^2S^2D_z^2- t_0^2 (S \kappa)^2 $ .

Finally, 
\begin{align}
    E^\pm(K)%&=\frac{1}{\sqrt{2}} \left( -\Tilde{A} \pm \sqrt{\Tilde{A}^2 - 4 \Tilde{B}} \right)^{\frac{1}{2}}\\
    %&= \frac{1}{\sqrt{2}} \left( 2t_0^2 - 2S^2 D_z^2 -  (S K)^2 \pm \sqrt{S^2D_z^2(S K)^2 + (S K)^4} \right)^{\frac{1}{2}} \\
    = \frac{1}{\sqrt{2}} \sqrt{ 2t_0^2 - 2S^2 D_z^2 -  (S \kappa)^2 \pm  S^2\kappa\sqrt{D_z^2 + \kappa^2}}
    \approx \frac{1}{\sqrt{2}} \sqrt{ 2t_0^2 - 2S^2 D_z^2 -  (S \kappa)^2 \pm (S\kappa)^2}
\end{align}
where in the last step, we simplified $\sqrt{D_z^2+\kappa^2} \approx \kappa$ since in our parameter set $D_z\ll \kappa$. So the gap can be approximated by $E_g \approx \sqrt{t_0^2 - S^2 D_z^2}-\sqrt{t_0^2 - S^2 D_z^2-(S\kappa)^2}$,

which for our parameter set gives \SI{2.7}{meV}.

\section{Numerical calculation of dispersion relation}\label{E:NumericalCalculationRD}

Here we presented the details of obtaining dispersion relation from the numerical spin dynamics calculation.

In the first step, we calculate the difference of the spin direction in a given time-step with respect to spin direction at the time-step $t=0$.

\begin{equation}
\hat{\textbf{S}}^{\alpha}\left( \textbf{r},t\right)=\textbf{S}^{\alpha}\left( \textbf{r},t\right)-\textbf{S}^{\alpha}\left( \textbf{r},t=0\right)
\end{equation}

where $\textbf{r}$ is the position of the spin and $\alpha=x,y,z$ are spin direction components. 

Then using Python Numpy FFT libraries the spin position-frequency domain was calculated for each spin over all time steps $T$:

\begin{equation}
\hat{\textbf{S}}^{\alpha}\left( \textbf{r},\omega\right)=\sum_{t=0}^{T}\hat{\textbf{S}}^{\alpha}\left( \textbf{r},t\right)e^{-i\omega t},
\end{equation}

In order to find Fourier transform in the wavevector domain, we implemented the summation over all spins in the system for a particular wavevector $\textbf{k}=k^1\hat{b}_1+k^2\hat{b}_2$, where $k^1$ and $k^2$ are wavevector components in reciprocal basis $\hat{b}_1, \hat{b}_2$.

\begin{equation}
\hat{\textbf{S}}^{\alpha}\left( \textbf{k},\omega \right)=\sum_{m}^{M}\sum_{n}^{N}\sum_{p}^{P=2}\hat{\textbf{S}}^{\alpha}\left( \textbf{r}_{m,n}^{p},\omega\right)e^{-i\textbf{k}\cdot(\textbf{r}_{m,n}-\textbf{r}_{m,n}^{p})}e^{-i\left(k^1m+k^2n\right)},
\end{equation}

where $m$ and $n$ are indices of the unit cell in the basis of the honeycomb lattice, and $M$ and $N$ is the number of unit cells in each direction. The particular spin in the unit cell is denoted as $p$, $\textbf{r}_{m,n}$ is the position of a unit cell for a node $n$,$m$ and $\textbf{r}_{m,n}^{p}$ is the position of the $p$-th spin in the unit cell.

\twocolumngrid

\bibliography{library}

\clearpage

\end{document}